\documentclass[aps,reprint,superscriptaddress]{revtex4-2}
\pdfoutput=1
\usepackage{braket,dsfont,subfigure,amsfonts,amssymb,bm,graphicx,float,amsmath,hyperref,natbib}
\usepackage{soul}
\usepackage[utf8]{inputenc}
\usepackage[english]{babel}
\usepackage[T1]{fontenc}
\usepackage{tikz}
\usepackage{lipsum}

\newcommand{\sect}[1]{\bigbreak\paragraph*{#1.---}}

\newcommand{\dif}{\mathop{ }\!\mathrm{d}}
\newcommand{\romnum}[1]{\MakeUppercase{\romannumeral #1}}

\usepackage[linesnumbered,ruled,vlined]{algorithm2e}
\SetKwInput{KwInput}{Input}
\SetKwInput{KwOutput}{Output}
\SetKwRepeat{Do}{do}{while}
\let\oldnl\nl
\newcommand{\nonl}{\renewcommand{\nl}{\let\nl\oldnl}}

\begin{document}

\title{Variational Matrix Product State Approach for Non-Hermitian System Based on a Companion Hermitian Hamiltonian}

\author{Zhen Guo}
\affiliation{State Key Laboratory of Low Dimensional Quantum Physics, Department of Physics, Tsinghua University, Beijing 100084, China}
\author{Zheng-Tao Xu}
\affiliation{State Key Laboratory of Low Dimensional Quantum Physics, Department of Physics, Tsinghua University, Beijing 100084, China}
\author{Meng Li}
\affiliation{State Key Laboratory of Low Dimensional Quantum Physics, Department of Physics, Tsinghua University, Beijing 100084, China}
\author{Li You}
\email{lyou@tsinghua.edu.cn}
\affiliation{State Key Laboratory of Low Dimensional Quantum Physics, Department of Physics, Tsinghua University, Beijing 100084, China}
\affiliation{Frontier Science Center for Quantum Information, Beijing, China}
\affiliation{Hefei National Laboratory, Hefei, 230088, China}
\affiliation{Beijing Academy of Quantum Information Sciences, Beijing 100193, China}
\author{Shuo Yang}
\email{shuoyang@tsinghua.edu.cn}
\affiliation{State Key Laboratory of Low Dimensional Quantum Physics, Department of Physics, Tsinghua University, Beijing 100084, China}
\affiliation{Frontier Science Center for Quantum Information, Beijing, China}
\affiliation{Hefei National Laboratory, Hefei, 230088, China}

\begin{abstract}
Non-Hermitian systems exhibiting topological properties are attracting growing interest. 
In this work, we propose an algorithm for solving the ground state of a non-Hermitian system in the matrix product state (MPS) formalism based on a companion Hermitian Hamiltonian.
If the eigenvalues of the non-Hermitian system are known, the companion Hermitian Hamiltonian can be directly constructed and solved using Hermitian variational methods.
When the eigenvalues are unknown, a gradient descent along with the companion Hermitian Hamiltonian yields both the ground state eigenenergy and the eigenstate.
With the variational principle as a solid foundation, our algorithm ensures convergence and provides results in excellent agreement with the exact solutions of the non-Hermitian Su-Schrieffer-Heeger (nH-SSH) model as well as its interacting extension. 
The approach we present avoids solving any non-Hermitian matrix and overcomes numerical instabilities commonly encountered in large non-Hermitian systems.
\end{abstract}
\maketitle

\sect{Introduction}

The idea of using non-Hermitian Hamiltonian to effectively describe open system backdates to the mid-1900s~\cite{Gamow1928,Feshbach1954,Feshbach1964}, not long after the birth of Quantum Mechanics. 
Over the years, non-Hermitian Hamiltonians have arisen in a variety of non-conservative systems, both classical~\cite{Schindler2011,Bender2013,Bittner2012,Fu2020,Jin2018,Makris2008,Tang2022} and quantum~\cite{Kreibich2014,Xu2015,Kepesidis2016,Lee2014,Zhang2020a,Zhang2021,Ling2022,Hashimoto2015,Roccati2022}. 
In systems exhibiting non-Hermitian skin effects, the conventional bulk-boundary correspondence (BBC) is broken~\cite{Yao2018,Li2020,Alsallom2021,Zhang2020b,Song2019,Lee2020,Yi2020}.
Significant efforts are being made to characterize non-Hermitian BBC, such as defining BBC through singular value gap~\cite{Herviou2019a}, detecting BBC using entanglement entropy~\cite{Chen2022,Chang2020,Herviou2019}, and understanding BBC by generalized Bloch theory~\cite{Yokomizo2019,Yang2020a,Yokomizo2020}.

Most studies of non-Hermitian systems focus on single-particle Hamiltonians, but not all properties of non-Hermitian many-body states can be directly derived from single-particle wave functions, even in the non-interacting case~\cite{Alsallom2021}. 
Beyond the single-particle research, standard methods for many-body non-Hermitian systems are often limited to small system size \cite{Sun2022,Chen2022,Hamazaki2019,Lee2020} due to the exponential growth of Hilbert space. 
Fortunately, not all quantum states in the many-body Hilbert space are equally important to the main physics. 
For Hermitian systems, low energy states of realistic Hamiltonians are constrained by locality and obey the entanglement area law~\cite{Verstraete2006,Hastings2007}. 
Tensor network (TN) states can be constructed based on this property, allowing them to naturally capture the most relevant states in Hilbert space.
In recent years, TN has emerged as a powerful tool to study strongly correlated quantum many-body systems. 
When solving the ground state of a one-dimensional (1D) Hermitian Hamiltonian, the variational matrix product state (VMPS) method~\cite{Verstraete2004, Schollwoeck2011} and the equivalent density matrix renormalization group (DMRG) algorithm~\cite{White1992} always converge, as guaranteed by the variational principle.
For large strongly correlated non-Hermitian systems, analogous principles and stable algorithms are desired.

Although DMRG has been successfully applied to certain non-Hermitian problems~\cite{Carlon1999,Hieida1998,Rotureau2006,Yamamoto2022,Zhang2020c}, debates remain regarding the choices of density matrices~\cite{Wang1997,Enss2001,Huang2011a,Carlon1999,Yamamoto2022,Peschel,Nishino1999,Zhang2020c,Chan2005} and numerical difficulties remain for some approaches.
For instance, Ref.~\cite{Chan2005} points out that the convergence characteristics of non-Hermitian algorithms are less favorable than in the Hermitian case, with the storage and computing time more than doubled due to the inequivalent left and right eigenspaces.
Reference~\cite{Rotureau2006} reports an instability of non-Hermitian DMRG even after many iterations and attributes this to non-Hermiticity.
References~\cite{Huang2011, Huang2011a, Zhang2020c} introduce biorthonormal DMRG approaches capable of providing accurate results while remaining problematic at exceptional points. 
Our study ascribes the latter to the biorthonormal condition itself since nearly orthogonal left and right eigenvectors are commonly found in non-Hermitian systems exhibiting skin effect~\cite{sm}.

In this work, we introduce two practical VMPS approaches for a non-Hermitian system based on a companion Hermitian Hamiltonian.
The Hermitian variational principle consequently guarantees convergence, and we avoid all numerical difficulties by not directly solving any non-Hermitian matrix.

\sect{Variational principle\label{sec:nh}}

The VMPS method, like other Hermitian variational methods, relies on the fact that a ground state has the lowest energy.
To go beyond this Hermitian paradigm, the first step is to define the ground state of a non-Hermitian system and its energy gradient.
Two common definitions of ground states are used, depending on whether the real or imaginary parts of energy are minimized, with the state denoted by $|{\rm SR}\rangle$ or $|{\rm SI}\rangle$, respectively.
For convenience, the ground state mentioned below refers to the right-eigenvector $|r \rangle$ unless otherwise specified, i.e., $H |r\rangle= e |r\rangle $.
To ensure numerical stability, we impose normalization conditions on the left and right eigenstates separately, such that $\langle l | l \rangle=1$ and $\langle r | r \rangle=1$.
A normalized right-eigenvector has the form $|r \rangle = |x\rangle / \sqrt{\langle x|x \rangle}$, and the corresponding expectation energy becomes $e(|x\rangle)=\langle x|H|x\rangle/\langle x|x\rangle$.
Because $e(|x\rangle)$ is not a holomorphic function, the derivative is not well-defined or cannot be used for straightforward optimization~\cite{UtrerasAlarcon2019}. 
Nevertheless, if the real and imaginary parts of the energy are treated as two real-valued functions of complex variables, the Wirtinger derivative~\cite{Wirtinger1927} $\partial_{\langle x|} \triangleq \frac12(\partial_{\Re\{|x\rangle\}} + i\partial_{\Im\{|x\rangle\}})$ can be adopted instead~\cite{sm}:
\begin{equation}
\partial_{\langle x|}\Re\{e(|x\rangle)\}= \frac{(H+H^\dag )|x\rangle}{2\langle x|x\rangle} - \frac{\langle x|(H+H^\dag)|x\rangle|x\rangle}{2\langle x|x\rangle^2}.
\label{equ:re}
\end{equation}

Naively, one expects to use Eq.~(\ref{equ:re}) for gradient descent to find the $| \rm SR \rangle$ ground state of a non-Hermitian system. 
The gradient $\partial_{\langle x|}\Re\{e(|x\rangle)\}$ should become zero at the end of iterations, resulting in $(H+H^\dag)|x\rangle\propto |x\rangle$.
However, such a condition shows that the converged $|x\rangle$ is an eigenvector of $H+H^\dag$ rather than $H$.
The states consequently obtained are usually not eigenstates of the non-Hermitian Hamiltonian $H$.
Therefore, the conventional variational principle is no longer directly suitable for non-Hermitian systems, unless a proper cost function is available.

\sect{Algorithm\label{sec:alg}}

We propose to take the eigenvector residual norm
\begin{equation}
\begin{aligned}
\mathcal{N}(|x\rangle) & \triangleq\left|H|x\rangle-e(|x\rangle)|x\rangle\right|^2
\end{aligned}
\label{equ:g}
\end{equation}
as the cost function. 
It is worth noting that any eigenstate of $H$, not just the ground state, fulfills $\mathcal{N}(|x\rangle) = 0$, and $\mathcal{N}(|x\rangle)$ is always real and non-negative. 
The Wirtinger derivative of $\mathcal{N}(|x\rangle)$ is simplified to~\cite{sm}
\begin{equation}
\begin{aligned}
\partial_{\langle x|}\mathcal{N}(|x\rangle) = & [H^\dag -e^{*}(|x\rangle)][H-e(|x\rangle)]|x\rangle \\
\triangleq & \mathbb{G}(H,e(|x\rangle)) |x\rangle.
\end{aligned}
\label{equ:dg}
\end{equation}
Our goal is then reduced to finding the lowest-energy state satisfying $\mathbb{G}(H,e(|x\rangle))|x\rangle=0$.

Before attempting to solve the above equation, we first establish a relation between the companion Hermitian Hamiltonian $\mathbb{G}(H,\varepsilon)$ and the original non-Hermitian Hamiltonian $H$.
Here $\mathbb{G}(H,\varepsilon)$ is non-negative definite and Hermitian for any arbitrary $\varepsilon$.
Using singular value decomposition
\begin{equation}
H-\varepsilon=USV^\dag,\;U^\dag U=V^\dag V=I,
\label{equ:svd}
\end{equation}
$\mathbb{G}(H,\varepsilon)$ is decomposed to
\begin{equation}
\mathbb{G}(H,\varepsilon)=(H^\dag-\varepsilon^*)(H-\varepsilon) = VS^2V^\dag,
\label{equ:gsvd}
\end{equation}
where singular values $s_i$ in $S$ are sorted in descending order.
We can see that $\mathbb{G}(H,\varepsilon)$ has the smallest eigenvalue $s_n^2=0$ if and only if $H$ has an eigenvalue $\varepsilon$.
Furthermore, the vector $V_n$ is a shared eigenstate of $H$ and $\mathbb{G}(H,\varepsilon)$, with eigenvalues $\varepsilon$ and $s_n^2=0$, respectively.

For non-interacting systems, the energy $e$ of a many-body ground state can be obtained from summing up single-particle energies. 
Finding the ground state of a non-Hermitian Hamiltonian $H$ therefore reduces to finding the zero-energy ground state of the companion Hermitian Hamiltonian $\mathbb{G}(H, e)$, for which the powerful VMPS approach can be employed. 
In practice, given a finite virtual bond dimension $D$ of MPS, the ground energy of $\mathbb{G}(H, e)$ after convergence will retain a tiny non-zero value $\eta$, and $s_n=\sqrt{\eta}$ measures whether $D$ is large enough.
Hereafter, we will refer to the aforementioned algorithm with supplied eigenenergies as the Hermitianized variational matrix product state (HVMPS) method, which is guaranteed to converge according to the standard VMPS method.

When eigenvalues are not provided or unknown, a gradient descent method facilitated by the companion Hermitian Hamiltonian can determine the ground energy and the corresponding eigenstate simultaneously, and this will be called the gradient variational matrix product state (GVMPS) method.
For simplicity, we illustrate GVMPS using a parity-time ($\mathcal{PT}$) symmetric system since its many-body ground energy is always real.
More details and its application to non-$\mathcal{PT}$ symmetric models are given in Supplemental Material.

The Wirtinger derivative of $s_n(\varepsilon)$ with respect to $\varepsilon$ reads~\cite{sm}
\begin{equation}
    \partial_{\varepsilon^*}s_n=\frac{\varepsilon-V_n^\dag H V_n}{2s_n}\label{equ:wd}.
\end{equation}
One may employ the gradient descent method to find the smallest $\varepsilon_{\mathrm{opt}}$ that minimizes $s_n(\varepsilon)$, such that $\varepsilon_{\mathrm{opt}}$ becomes the ground state energy of $H$ and the corresponding eigenvector $V_n$ is obtained at the same time.
However, when $s_n(\varepsilon)$ approaches its minimum, the gradient is usually close to zero, which slows down the convergence and even results in instabilities~\cite{sm}.
This can be avoided by manually setting the gradient to $\varepsilon-V_n^\dag H V_n$ and using an adaptive learning rate~\cite{sm}.
Regarding the initial value of $\varepsilon$, we note that
\begin{equation}
\Re\left\{e(|x\rangle) \right\} = \frac{\langle x|(H+H^\dag)|x\rangle}{2\langle x|x\rangle} \ge \tau,
\end{equation} 
where $\tau$ is the smallest eigenvalue of the Hermitian matrix $(H+H^\dag)/2$.
Therefore, all eigenvalues of $H$ have real parts greater than $\tau$, making $\tau$ a good starting point for determining the eigenvalue of $|\rm SR \rangle$.
Since the gradient descent only depends on one scalar parameter $\varepsilon$ and there is no other local minimum from the initial value $\tau$ to the desired energy, the GVMPS method is found to be efficient and well-converged.

Both the HVMPS and GVMPS approaches we propose avoid directly solving non-Hermitian problems, and the companion Hermitian Hamiltonian helps to prevent typical numerical instabilities of non-Hermitian systems. 
Some of our benchmark results are presented below.

\sect{Non-interacting model\label{sec:results}}

We first test for non-interacting systems using the 1D nH-SSH model, which exhibits interesting topological properties and has recently received a lot of attention~\cite{Yao2018, Lieu2018,Chang2020,Han2021,Xi2021}.
The Hamiltonian takes the following form
\begin{equation}
\begin{aligned}
H_0 = & \sum\limits_i \left[(t+\gamma/2)~a^\dag_i b_i + (t-\gamma/2)~b^\dag_i a_i\right. \\&\left.+b^\dag_i a_{i+1}+a_{i+1}^\dag b_i\right],
\end{aligned}
\end{equation} 
with one unit cell composed of two sites. 
The intra-cell hopping is non-reciprocal and characterized by a non-Hermitian strength $\gamma$, while the inter-cell hopping is Hermitian and set as unit strength. 
High-order exceptional points for this model~\cite{Fernandez2018,Chang2020,Heiss2012,Tzeng2021} appear at $|t|=|\gamma/2|$.

Since MPS methods are more accurate and efficient for finding low entanglement states that satisfy the area law, it is necessary to investigate the entanglement properties throughout the parameter space to determine which regions are more favourably described by MPS.
Under periodic boundary condition (PBC), previous study~\cite{Guo2021} on the bi-orthogonal entanglement entropy (EE)~\cite{Guo2021,Chang2020,Herviou2019,Tu2021,Brody2013,Pati2009} of the nH-SSH model finds its ground state obeys the area law at $t=1$ and $\gamma>4$.
Since our algorithms only focus on the right eigenstate and the bi-orthogonal condition may induce extra numerical difficulties~\cite{sm}, we will use the EE of the right eigenstate $|r \rangle$ instead. 
The bipartite EE between subsystem $A$ and its complementary part $\bar{A}$ is given by $S_A = -\mathrm{Tr}(\rho_A \mathrm{ln} \rho_A)$, where $\rho_A=\mathrm{Tr}_{\bar{A}} \rho^{rr}$, $\rho^{rr} = |r \rangle \langle r|/ \langle r | r \rangle$, and the length of $A$ is $L_A$.
Usually, $S_A$ reaches its maximum when $L_A$ is half of the total length, and we call it the maximum EE.
As shown in the Supplemental Material, this definition of EE reveals the same area-law behavior for $t=1$ and $\gamma>4$.

\begin{figure}[tbp]
\centering
\includegraphics[width=1.0\columnwidth]{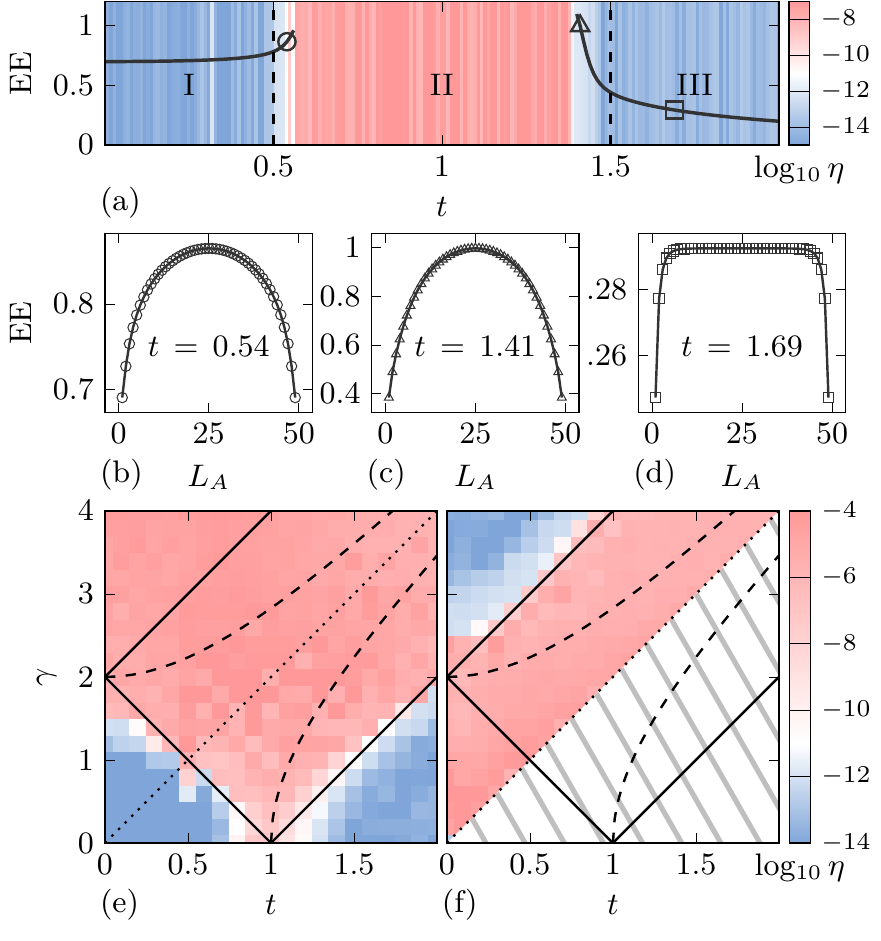}
\caption{
(Color online)
(a) Solid lines show the maximum EE of $|{\rm SR}\rangle$ calculated for the OBC nH-SSH model at $\gamma=1$ for various $t$. 
The missing region indicates no convergence found with bond dimension $D=300$. 
The background color represents the logarithm of the converged ground energy of $\mathbb{G}(H, \varepsilon)$, i.e., $\log_{10}\eta$.
(b-d) Bipartite EE as a function of the subsystem length $L_A$.
(e-f) $\log_{10}\eta$ in the $(t,\gamma)$ parameter space for the ground states $|{\rm SR}\rangle$ (e) and $|{\rm SI}\rangle$ (f), calculated with bond dimension $D=100$. 
The blue regions obey the area law.
Black dashed (solid) lines are the topological phase boundaries defined by energy gap closing points for OBC (PBC). 
Dotted lines indicate exceptional points, beneath which the energy spectra are real.
}
\label{fig:entropyobc}
\end{figure}

Now we apply HVMPS to investigate the entanglement behavior of the $|\mathrm{SR} \rangle$ ground state of the $25$-unit-cell nH-SSH model under open boundary condition (OBC), with the many-body energy supplied by the sum of single-particle energies. 
As shown in Fig.~\ref{fig:entropyobc}(a), we find three regions with different entropy distributions for $\gamma=1$, roughly separated by $t = 0.5$ and $1.5$. 
The solid lines in regions \romnum 1 and \romnum 3 display the maximum EE as a function of $t$, and their convergence is confirmed by increasing the virtual bond dimension $D$.
Typical EE in these regions as a function of $L_A$ is shown in Fig.~\ref{fig:entropyobc}(d), where the plateau in the middle region indicates area-law behavior. 
In most parts of region \romnum 2, a stable entropy distribution is not reached for $D=300$.
Near the boundaries of region \romnum 2, area-law violations are observed and shown in Figs.~\ref{fig:entropyobc}(b) and (c).
As indicated by the background color in Fig.~\ref{fig:entropyobc}(a), the logarithm of the converged ground energy $\eta$ of $\mathbb{G}(H, \varepsilon)$ in region \romnum 2 is a few orders of magnitude larger than in regions \romnum 1 and \romnum 3. 
In fact, a bond dimension $D \sim 100$ is enough to reach $\eta<10^{-13}$ in regions \romnum 1 and \romnum 3, whereas the required $D$ quickly exceeds $300$ once entering region \romnum 2.
Together with the sudden increase of the maximum EE near the boundaries shown in Fig.~\ref{fig:entropyobc}(a), we conclude that area law is violated in region \romnum 2.

Similarly, we sweep the entire parameter space for $|{\rm SR}\rangle$ and $|{\rm SI}\rangle$ ground states using $\log_{10}\eta$ as a criterion, and the area-law-obeyed regions are painted in blue in Figs.~\ref{fig:entropyobc}(e) and (f). 
Remarkably, the OBC area law boundaries coincide with the PBC topological phase boundaries, which are solid lines in Figs.~\ref{fig:entropyobc}(e) and (f). 
The BBC is broken for the nH-SSH model, and the energy gap closing points under OBC (dashed lines in Figs.~\ref{fig:entropyobc}(e) and (f)) can no longer be taken as good indicators for bulk phase boundaries~\cite{Herviou2019}. 
Nevertheless, the EE of ground states under OBC contains bulk phase information like in the Hermitian case, which helps to restore BBC.

According to Ref.~\cite{Korff2008}, a local non-Hermitian Hamiltonian with real spectra can be mapped to a non-local Hermitian one by a similar transformation, with area law no longer necessarily valid.
In our case, the ground state of a local non-Hermitian Hamiltonian $H$ becomes the ground state of the companion Hermitian Hamiltonian $(H^\dag-e^*)(H-e)$, possessing long range interactions. 
Our algorithm thus provides an insightful understanding for area law violation in a non-Hermitian system.

\begin{figure}[tbp]
\centering
\includegraphics[width=1.0\columnwidth]{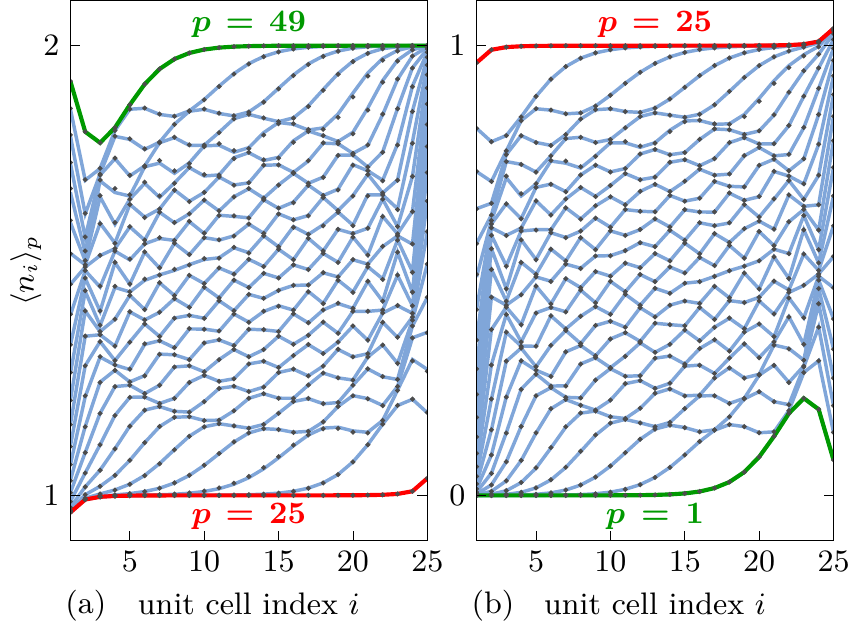}
\caption{
(Color online)
Particle distributions of the many-body $|{\rm SR}\rangle$ state for the $25$-unit-cell nH-SSH model at $t=1.8$ and $\gamma=1.3$ for different particle numbers (a) $p=25,26,\cdots,49$ and (b) $p=1,2,\cdots,25$.
Black dots denote the GVMPS results with $U(1)$-symmetry and $D=100$.
Solid lines are obtained based on single-particle wave functions. 
The green line labeled by $p=1$ ($p=49$) represents the single-particle (single-hole) $|\rm SR \rangle$ ground state, while the red lines labeled by $p=25$ denote the ground state at half-filling.
}
\label{fig:skin}
\end{figure}

In regions where area law is satisfied, the GVMPS algorithm can be applied to find both ground state energy and wave function. 
The accuracy of the wave function is benchmarked by calculating the many-body non-Hermitian skin effect arising from non-reciprocal hoppings.
In the pioneering work of Ref.~\cite{Yao2018}, a diagonal matrix is introduced to transform the single-particle nH-SSH model to a Hermitian one, with the diagonal elements of the transformation decaying quickly with cell index implicating skin effect, which can be clearly interpreted by a generalized Bloch theory with complex momentum~\cite{Yao2018, Yokomizo2019,Song2019}. 
In addition, spectral instability from nonnormal Hamiltonians can be used to interpret skin effect~\cite{Okuma2020}. 
While these discussions mostly address single-particle situations, the skin effect itself is thought to be greatly suppressed for many-body states due to Pauli exclusion~\cite{Alsallom2021, Lee2020}. 
However, most many-body studies on non-Hermitian systems employ exact diagonalization (ED) and are limited to small sizes, leaving larger systems insufficiently investigated.

By implementing $U(1)$-symmetry~\cite{Rakov2018} to the GVMPS algorithm, the many-body $| \mathrm{SR} \rangle$ ground state particle distributions are found and plotted for the $25$-unit-cell nH-SSH model.
The black dots in Fig.~\ref{fig:skin} show how particle distributions $\langle n_{i} \rangle_p = \langle n_{i}^{a} \rangle_p + \langle n_{i}^{b} \rangle_p $ evolve as the particle number $p$ increases, where $i$ is the unit-cell index, $n_{i}^{a} = a_{i}^{\dagger}a_{i}$, and $n_{i}^{b} = b_{i}^{\dagger}b_{i}$.
For comparison, we also derive a precise and efficient method~\cite{sm} to calculate particle distributions from single-particle wave functions, as shown by the solid lines in Fig.~\ref{fig:skin} and in excellent agreement with GVMPS results.
The red lines in Fig.~\ref{fig:skin} denote the distribution at half filling, where skin effect only causes a slight upturn. 
Apart from adding the first and last particles when single-particle skin effects are clearly revealed, the many-body particle distributions exhibit inhomogeneous density waves.
Moreover, distributions with $p>25$ are centro-symmetric with respect to those of $p\le25$, i.e., $\langle n_{i}\rangle_p = 2-\langle n_{N+1-i}\rangle_{2N-p}$ for the $N$-unit-cell system~\cite{sm}. 

\sect{Interacting model} 
For interacting non-Hermitian systems, the results from our GVMPS are compared with the ones from ED.
We calculate the ground state of a $12$-unit-cell OBC nH-SSH model with nearest-neighbor repulsions at $t=0.5$ and $\gamma=1$.
The Hamiltonian is given by
\begin{equation}
H_1 = H_0+ \sum\limits_i U \left(n_{i}^{a} n_{i}^{b} +  n_{i}^{b} n_{i+1}^{a} \right).
\end{equation}

\begin{figure}[tbp]
\centering
\includegraphics[width=1.0\columnwidth]{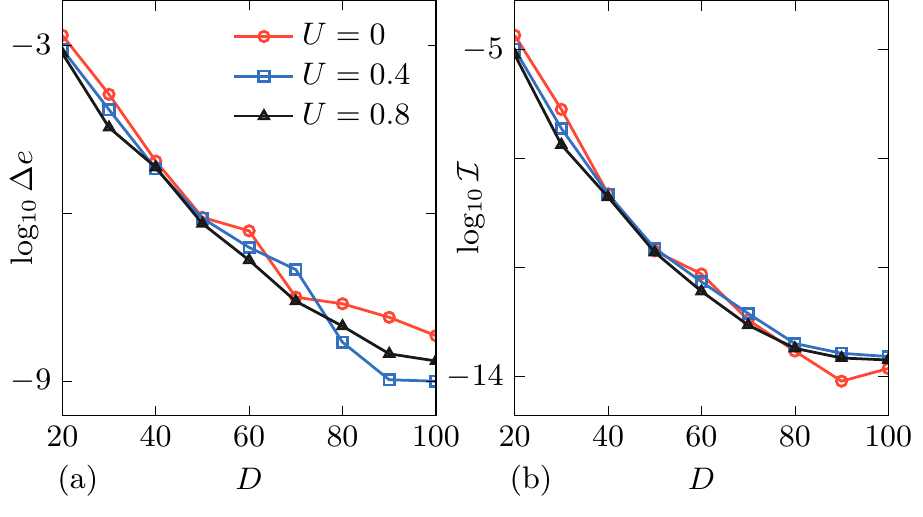}
\caption{
(Color online)
The $D$-dependence of (a) energy error $\Delta e$ and (b) infidelity $\mathcal{I}$ for the $12$-unit-cell nH-SSH model with different $U$ at $t=0.5$ and $\gamma=1$.
}
\label{fig:ed}
\end{figure}

The energy error $\Delta e = |e_{\rm GVMPS}-e_{\rm ED}|$ and wave function infidelity $\mathcal{I}=\left| 1-|\langle\psi_{\rm ED}|\psi_{\rm GVMPS}\rangle|^2 \right|$ of the two methods are shown in Fig.~\ref{fig:ed}.
Both quantities decrease rapidly as the bond dimension $D$ increases, and sufficiently accurate results are obtained with $D$ less than $100$.
With the help of the companion Hermitian Hamiltonian, we are not required to solve any non-Hermitian matrix and no instability is observed even at $U=0$, where high-order exceptional points are predicted to emerge.

\sect{Conclusion}
We propose two efficient MPS methods for non-Hermitian systems.
If the eigenenergies of a non-Hermitian Hamiltonian are known, HVMPS can be employed to find its eigenstates by solving a companion Hermitian Hamiltonian.
When the eigenenergies are unknown, HVMPS can be combined with gradient descent and upgraded to GVMPS, such that both ground energy and wave function can be obtained.
These methods provide accurate results for non-interacting and interacting non-Hermitian systems.
Since no non-Hermitian matrix is directly solved, our approach mitigates all known numerical difficulties of typical non-Hermitian systems and ensures convergence through the variational principle. 
The algorithms presented here are robust for finding many-body wave functions of a general Hamiltonian, which could pave the way for studying large-scale strongly correlated non-Hermitian systems.

\begin{acknowledgments}
This work is supported by the National Key R\&D Program of China (Grant No. 2018YFA0306504), the National Natural Science Foundation of China (NSFC) (Grants Nos. 12174214, 92065205, 11654001, and U1930201), and by Innovation Program for Quantum Science and Technology (Project 2-9-4).
\end{acknowledgments}

\bibliography{NHMps}

\begin{thebibliography}{69}%
\makeatletter
\providecommand \@ifxundefined [1]{%
 \@ifx{#1\undefined}
}%
\providecommand \@ifnum [1]{%
 \ifnum #1\expandafter \@firstoftwo
 \else \expandafter \@secondoftwo
 \fi
}%
\providecommand \@ifx [1]{%
 \ifx #1\expandafter \@firstoftwo
 \else \expandafter \@secondoftwo
 \fi
}%
\providecommand \natexlab [1]{#1}%
\providecommand \enquote  [1]{``#1''}%
\providecommand \bibnamefont  [1]{#1}%
\providecommand \bibfnamefont [1]{#1}%
\providecommand \citenamefont [1]{#1}%
\providecommand \href@noop [0]{\@secondoftwo}%
\providecommand \href [0]{\begingroup \@sanitize@url \@href}%
\providecommand \@href[1]{\@@startlink{#1}\@@href}%
\providecommand \@@href[1]{\endgroup#1\@@endlink}%
\providecommand \@sanitize@url [0]{\catcode `\\12\catcode `\$12\catcode
  `\&12\catcode `\#12\catcode `\^12\catcode `\_12\catcode `\%12\relax}%
\providecommand \@@startlink[1]{}%
\providecommand \@@endlink[0]{}%
\providecommand \url  [0]{\begingroup\@sanitize@url \@url }%
\providecommand \@url [1]{\endgroup\@href {#1}{\urlprefix }}%
\providecommand \urlprefix  [0]{URL }%
\providecommand \Eprint [0]{\href }%
\providecommand \doibase [0]{https://doi.org/}%
\providecommand \selectlanguage [0]{\@gobble}%
\providecommand \bibinfo  [0]{\@secondoftwo}%
\providecommand \bibfield  [0]{\@secondoftwo}%
\providecommand \translation [1]{[#1]}%
\providecommand \BibitemOpen [0]{}%
\providecommand \bibitemStop [0]{}%
\providecommand \bibitemNoStop [0]{.\EOS\space}%
\providecommand \EOS [0]{\spacefactor3000\relax}%
\providecommand \BibitemShut  [1]{\csname bibitem#1\endcsname}%
\let\auto@bib@innerbib\@empty
\bibitem [{\citenamefont {Gamow}(1928)}]{Gamow1928}%
  \BibitemOpen
  \bibfield  {author} {\bibinfo {author} {\bibfnamefont {G.}~\bibnamefont
  {Gamow}},\ }\bibfield  {title} {\bibinfo {title} {Zur quantentheorie des
  atomkernes},\ }\href {https://doi.org/10.1007/BF01343196} {\bibfield
  {journal} {\bibinfo  {journal} {Z. Angew. Phys.}\ }\textbf {\bibinfo {volume}
  {51}},\ \bibinfo {pages} {204} (\bibinfo {year} {1928})}\BibitemShut
  {NoStop}%
\bibitem [{\citenamefont {Feshbach}\ \emph {et~al.}(1954)\citenamefont
  {Feshbach}, \citenamefont {Porter},\ and\ \citenamefont
  {Weisskopf}}]{Feshbach1954}%
  \BibitemOpen
  \bibfield  {author} {\bibinfo {author} {\bibfnamefont {H.}~\bibnamefont
  {Feshbach}}, \bibinfo {author} {\bibfnamefont {C.~E.}\ \bibnamefont
  {Porter}},\ and\ \bibinfo {author} {\bibfnamefont {V.~F.}\ \bibnamefont
  {Weisskopf}},\ }\bibfield  {title} {\bibinfo {title} {Model for nuclear
  reactions with neutrons},\ }\href {https://doi.org/10.1103/PhysRev.96.448}
  {\bibfield  {journal} {\bibinfo  {journal} {Phys. Rev.}\ }\textbf {\bibinfo
  {volume} {96}},\ \bibinfo {pages} {448} (\bibinfo {year} {1954})}\BibitemShut
  {NoStop}%
\bibitem [{\citenamefont {Feshbach}(1964)}]{Feshbach1964}%
  \BibitemOpen
  \bibfield  {author} {\bibinfo {author} {\bibfnamefont {H.}~\bibnamefont
  {Feshbach}},\ }\bibfield  {title} {\bibinfo {title} {Unified theory of
  nuclear reactions},\ }\href {https://doi.org/10.1103/RevModPhys.36.1076}
  {\bibfield  {journal} {\bibinfo  {journal} {Rev. Mod. Phys.}\ }\textbf
  {\bibinfo {volume} {36}},\ \bibinfo {pages} {1076} (\bibinfo {year}
  {1964})}\BibitemShut {NoStop}%
\bibitem [{\citenamefont {Schindler}\ \emph {et~al.}(2011)\citenamefont
  {Schindler}, \citenamefont {Li}, \citenamefont {Zheng}, \citenamefont
  {Ellis},\ and\ \citenamefont {Kottos}}]{Schindler2011}%
  \BibitemOpen
  \bibfield  {author} {\bibinfo {author} {\bibfnamefont {J.}~\bibnamefont
  {Schindler}}, \bibinfo {author} {\bibfnamefont {A.}~\bibnamefont {Li}},
  \bibinfo {author} {\bibfnamefont {M.~C.}\ \bibnamefont {Zheng}}, \bibinfo
  {author} {\bibfnamefont {F.~M.}\ \bibnamefont {Ellis}},\ and\ \bibinfo
  {author} {\bibfnamefont {T.}~\bibnamefont {Kottos}},\ }\bibfield  {title}
  {\bibinfo {title} {Experimental study of active {LRC} circuits with
  $\mathcal{PT}$ symmetries},\ }\href
  {https://doi.org/10.1103/PhysRevA.84.040101} {\bibfield  {journal} {\bibinfo
  {journal} {Phys. Rev. A}\ }\textbf {\bibinfo {volume} {84}},\ \bibinfo
  {pages} {040101} (\bibinfo {year} {2011})}\BibitemShut {NoStop}%
\bibitem [{\citenamefont {Bender}\ \emph {et~al.}(2013)\citenamefont {Bender},
  \citenamefont {Berntson}, \citenamefont {Parker},\ and\ \citenamefont
  {Samuel}}]{Bender2013}%
  \BibitemOpen
  \bibfield  {author} {\bibinfo {author} {\bibfnamefont {C.~M.}\ \bibnamefont
  {Bender}}, \bibinfo {author} {\bibfnamefont {B.~K.}\ \bibnamefont
  {Berntson}}, \bibinfo {author} {\bibfnamefont {D.}~\bibnamefont {Parker}},\
  and\ \bibinfo {author} {\bibfnamefont {E.}~\bibnamefont {Samuel}},\
  }\bibfield  {title} {\bibinfo {title} {Observation of {PT} phase transition
  in a simple mechanical system},\ }\href {https://doi.org/10.1119/1.4789549}
  {\bibfield  {journal} {\bibinfo  {journal} {Am. J. Phys.}\ }\textbf {\bibinfo
  {volume} {81}},\ \bibinfo {pages} {173} (\bibinfo {year} {2013})}\BibitemShut
  {NoStop}%
\bibitem [{\citenamefont {Bittner}\ \emph {et~al.}(2012)\citenamefont
  {Bittner}, \citenamefont {Dietz}, \citenamefont {G{\"{u}}nther},
  \citenamefont {Harney}, \citenamefont {Miski-Oglu}, \citenamefont {Richter},\
  and\ \citenamefont {Sch{\"{a}}fer}}]{Bittner2012}%
  \BibitemOpen
  \bibfield  {author} {\bibinfo {author} {\bibfnamefont {S.}~\bibnamefont
  {Bittner}}, \bibinfo {author} {\bibfnamefont {B.}~\bibnamefont {Dietz}},
  \bibinfo {author} {\bibfnamefont {U.}~\bibnamefont {G{\"{u}}nther}}, \bibinfo
  {author} {\bibfnamefont {H.~L.}\ \bibnamefont {Harney}}, \bibinfo {author}
  {\bibfnamefont {M.}~\bibnamefont {Miski-Oglu}}, \bibinfo {author}
  {\bibfnamefont {A.}~\bibnamefont {Richter}},\ and\ \bibinfo {author}
  {\bibfnamefont {F.}~\bibnamefont {Sch{\"{a}}fer}},\ }\bibfield  {title}
  {\bibinfo {title} {$\mathcal{PT}$ symmetry and spontaneous symmetry breaking
  in a microwave billiard},\ }\href
  {https://doi.org/10.1103/PhysRevLett.108.024101} {\bibfield  {journal}
  {\bibinfo  {journal} {Phys. Rev. Lett.}\ }\textbf {\bibinfo {volume} {108}},\
  \bibinfo {pages} {024101} (\bibinfo {year} {2012})}\BibitemShut {NoStop}%
\bibitem [{\citenamefont {Fu}\ \emph {et~al.}(2020)\citenamefont {Fu},
  \citenamefont {Fu}, \citenamefont {Zhang}, \citenamefont {Wang},
  \citenamefont {Zhao},\ and\ \citenamefont {Ke}}]{Fu2020}%
  \BibitemOpen
  \bibfield  {author} {\bibinfo {author} {\bibfnamefont {Z.}~\bibnamefont
  {Fu}}, \bibinfo {author} {\bibfnamefont {N.}~\bibnamefont {Fu}}, \bibinfo
  {author} {\bibfnamefont {H.}~\bibnamefont {Zhang}}, \bibinfo {author}
  {\bibfnamefont {Z.}~\bibnamefont {Wang}}, \bibinfo {author} {\bibfnamefont
  {D.}~\bibnamefont {Zhao}},\ and\ \bibinfo {author} {\bibfnamefont
  {S.}~\bibnamefont {Ke}},\ }\bibfield  {title} {\bibinfo {title} {Extended
  {SSH} model in non-{Hermitian} waveguides with alternating real and imaginary
  couplings},\ }\href {https://doi.org/10.3390/app10103425} {\bibfield
  {journal} {\bibinfo  {journal} {Appl. Sci.}\ }\textbf {\bibinfo {volume}
  {10}},\ \bibinfo {pages} {3425} (\bibinfo {year} {2020})}\BibitemShut
  {NoStop}%
\bibitem [{\citenamefont {Jin}\ and\ \citenamefont {Song}(2018)}]{Jin2018}%
  \BibitemOpen
  \bibfield  {author} {\bibinfo {author} {\bibfnamefont {L.}~\bibnamefont
  {Jin}}\ and\ \bibinfo {author} {\bibfnamefont {Z.}~\bibnamefont {Song}},\
  }\bibfield  {title} {\bibinfo {title} {Incident direction independent wave
  propagation and unidirectional lasing},\ }\href
  {https://doi.org/10.1103/PhysRevLett.121.073901} {\bibfield  {journal}
  {\bibinfo  {journal} {Phys. Rev. Lett.}\ }\textbf {\bibinfo {volume} {121}},\
  \bibinfo {pages} {073901} (\bibinfo {year} {2018})}\BibitemShut {NoStop}%
\bibitem [{\citenamefont {Makris}\ \emph {et~al.}(2008)\citenamefont {Makris},
  \citenamefont {El-Ganainy}, \citenamefont {Christodoulides},\ and\
  \citenamefont {Musslimani}}]{Makris2008}%
  \BibitemOpen
  \bibfield  {author} {\bibinfo {author} {\bibfnamefont {K.~G.}\ \bibnamefont
  {Makris}}, \bibinfo {author} {\bibfnamefont {R.}~\bibnamefont {El-Ganainy}},
  \bibinfo {author} {\bibfnamefont {D.~N.}\ \bibnamefont {Christodoulides}},\
  and\ \bibinfo {author} {\bibfnamefont {Z.~H.}\ \bibnamefont {Musslimani}},\
  }\bibfield  {title} {\bibinfo {title} {Beam dynamics in
  $\mathcal{P}\mathcal{T}$ symmetric optical lattices},\ }\href
  {https://doi.org/10.1103/PhysRevLett.100.103904} {\bibfield  {journal}
  {\bibinfo  {journal} {Phys. Rev. Lett.}\ }\textbf {\bibinfo {volume} {100}},\
  \bibinfo {pages} {103904} (\bibinfo {year} {2008})}\BibitemShut {NoStop}%
\bibitem [{\citenamefont {Tang}\ \emph {et~al.}(2022)\citenamefont {Tang},
  \citenamefont {Ding},\ and\ \citenamefont {Ma}}]{Tang2022}%
  \BibitemOpen
  \bibfield  {author} {\bibinfo {author} {\bibfnamefont {W.}~\bibnamefont
  {Tang}}, \bibinfo {author} {\bibfnamefont {K.}~\bibnamefont {Ding}},\ and\
  \bibinfo {author} {\bibfnamefont {G.}~\bibnamefont {Ma}},\ }\bibfield
  {title} {\bibinfo {title} {Experimental realization of non-{Abelian}
  permutations in a three-state non-{Hermitian} system},\ }\href
  {https://doi.org/10.1093/NSR/NWAC010} {\bibfield  {journal} {\bibinfo
  {journal} {Natl. Sci. Rev.}\ ,\ \bibinfo {pages} {nwac010}} (\bibinfo {year}
  {2022})}\BibitemShut {NoStop}%
\bibitem [{\citenamefont {Kreibich}\ \emph {et~al.}(2014)\citenamefont
  {Kreibich}, \citenamefont {Main}, \citenamefont {Cartarius},\ and\
  \citenamefont {Wunner}}]{Kreibich2014}%
  \BibitemOpen
  \bibfield  {author} {\bibinfo {author} {\bibfnamefont {M.}~\bibnamefont
  {Kreibich}}, \bibinfo {author} {\bibfnamefont {J.}~\bibnamefont {Main}},
  \bibinfo {author} {\bibfnamefont {H.}~\bibnamefont {Cartarius}},\ and\
  \bibinfo {author} {\bibfnamefont {G.}~\bibnamefont {Wunner}},\ }\bibfield
  {title} {\bibinfo {title} {Realizing $\mathcal{PT}$-symmetric
  non-{Hermiticity} with ultracold atoms and {Hermitian} multiwell
  potentials},\ }\href {https://doi.org/10.1103/PhysRevA.90.033630} {\bibfield
  {journal} {\bibinfo  {journal} {Phys. Rev. A}\ }\textbf {\bibinfo {volume}
  {90}},\ \bibinfo {pages} {033630} (\bibinfo {year} {2014})}\BibitemShut
  {NoStop}%
\bibitem [{\citenamefont {Xu}\ \emph {et~al.}(2015)\citenamefont {Xu},
  \citenamefont {Liu}, \citenamefont {Sun},\ and\ \citenamefont {Li}}]{Xu2015}%
  \BibitemOpen
  \bibfield  {author} {\bibinfo {author} {\bibfnamefont {X.-W.}\ \bibnamefont
  {Xu}}, \bibinfo {author} {\bibfnamefont {Y.-x.}\ \bibnamefont {Liu}},
  \bibinfo {author} {\bibfnamefont {C.-P.}\ \bibnamefont {Sun}},\ and\ \bibinfo
  {author} {\bibfnamefont {Y.}~\bibnamefont {Li}},\ }\bibfield  {title}
  {\bibinfo {title} {Mechanical $\mathcal{PT}$ symmetry in coupled
  optomechanical systems},\ }\href {https://doi.org/10.1103/PhysRevA.92.013852}
  {\bibfield  {journal} {\bibinfo  {journal} {Phys. Rev. A}\ }\textbf {\bibinfo
  {volume} {92}},\ \bibinfo {pages} {013852} (\bibinfo {year}
  {2015})}\BibitemShut {NoStop}%
\bibitem [{\citenamefont {Kepesidis}\ \emph {et~al.}(2016)\citenamefont
  {Kepesidis}, \citenamefont {Milburn}, \citenamefont {Huber}, \citenamefont
  {Makris}, \citenamefont {Rotter},\ and\ \citenamefont
  {Rabl}}]{Kepesidis2016}%
  \BibitemOpen
  \bibfield  {author} {\bibinfo {author} {\bibfnamefont {K.~V.}\ \bibnamefont
  {Kepesidis}}, \bibinfo {author} {\bibfnamefont {T.~J.}\ \bibnamefont
  {Milburn}}, \bibinfo {author} {\bibfnamefont {J.}~\bibnamefont {Huber}},
  \bibinfo {author} {\bibfnamefont {K.~G.}\ \bibnamefont {Makris}}, \bibinfo
  {author} {\bibfnamefont {S.}~\bibnamefont {Rotter}},\ and\ \bibinfo {author}
  {\bibfnamefont {P.}~\bibnamefont {Rabl}},\ }\bibfield  {title} {\bibinfo
  {title} {$\mathcal{P}\mathcal{T}$-symmetry breaking in the steady state of
  microscopic gain-loss systems},\ }\href
  {https://doi.org/10.1088/1367-2630/18/9/095003} {\bibfield  {journal}
  {\bibinfo  {journal} {New J. Phys.}\ }\textbf {\bibinfo {volume} {18}},\
  \bibinfo {pages} {095003} (\bibinfo {year} {2016})}\BibitemShut {NoStop}%
\bibitem [{\citenamefont {Lee}\ \emph {et~al.}(2014)\citenamefont {Lee},
  \citenamefont {Reiter},\ and\ \citenamefont {Moiseyev}}]{Lee2014}%
  \BibitemOpen
  \bibfield  {author} {\bibinfo {author} {\bibfnamefont {T.~E.}\ \bibnamefont
  {Lee}}, \bibinfo {author} {\bibfnamefont {F.}~\bibnamefont {Reiter}},\ and\
  \bibinfo {author} {\bibfnamefont {N.}~\bibnamefont {Moiseyev}},\ }\bibfield
  {title} {\bibinfo {title} {Entanglement and spin squeezing in non-{Hermitian}
  phase transitions},\ }\href {https://doi.org/10.1103/PhysRevLett.113.250401}
  {\bibfield  {journal} {\bibinfo  {journal} {Phys. Rev. Lett.}\ }\textbf
  {\bibinfo {volume} {113}},\ \bibinfo {pages} {250401} (\bibinfo {year}
  {2014})}\BibitemShut {NoStop}%
\bibitem [{\citenamefont {Zhang}\ \emph
  {et~al.}(2020{\natexlab{a}})\citenamefont {Zhang}, \citenamefont {Jin},\ and\
  \citenamefont {Song}}]{Zhang2020a}%
  \BibitemOpen
  \bibfield  {author} {\bibinfo {author} {\bibfnamefont {X.~Z.}\ \bibnamefont
  {Zhang}}, \bibinfo {author} {\bibfnamefont {L.}~\bibnamefont {Jin}},\ and\
  \bibinfo {author} {\bibfnamefont {Z.}~\bibnamefont {Song}},\ }\bibfield
  {title} {\bibinfo {title} {Dynamic magnetization in non-{Hermitian} quantum
  spin systems},\ }\href {https://doi.org/10.1103/PhysRevB.101.224301}
  {\bibfield  {journal} {\bibinfo  {journal} {Phys. Rev. B}\ }\textbf {\bibinfo
  {volume} {101}},\ \bibinfo {pages} {224301} (\bibinfo {year}
  {2020}{\natexlab{a}})}\BibitemShut {NoStop}%
\bibitem [{\citenamefont {Zhang}\ \emph {et~al.}(2021)\citenamefont {Zhang},
  \citenamefont {Ouyang}, \citenamefont {Huang}, \citenamefont {Wang},
  \citenamefont {Zhang}, \citenamefont {Yu}, \citenamefont {Chang},
  \citenamefont {Liu}, \citenamefont {Deng},\ and\ \citenamefont
  {Duan}}]{Zhang2021}%
  \BibitemOpen
  \bibfield  {author} {\bibinfo {author} {\bibfnamefont {W.}~\bibnamefont
  {Zhang}}, \bibinfo {author} {\bibfnamefont {X.}~\bibnamefont {Ouyang}},
  \bibinfo {author} {\bibfnamefont {X.}~\bibnamefont {Huang}}, \bibinfo
  {author} {\bibfnamefont {X.}~\bibnamefont {Wang}}, \bibinfo {author}
  {\bibfnamefont {H.}~\bibnamefont {Zhang}}, \bibinfo {author} {\bibfnamefont
  {Y.}~\bibnamefont {Yu}}, \bibinfo {author} {\bibfnamefont {X.}~\bibnamefont
  {Chang}}, \bibinfo {author} {\bibfnamefont {Y.}~\bibnamefont {Liu}}, \bibinfo
  {author} {\bibfnamefont {D.-L.}\ \bibnamefont {Deng}},\ and\ \bibinfo
  {author} {\bibfnamefont {L.-M.}\ \bibnamefont {Duan}},\ }\bibfield  {title}
  {\bibinfo {title} {Observation of non-{Hermitian} topology with nonunitary
  dynamics of solid-state spins},\ }\href
  {https://doi.org/10.1103/PhysRevLett.127.090501} {\bibfield  {journal}
  {\bibinfo  {journal} {Phys. Rev. Lett.}\ }\textbf {\bibinfo {volume} {127}},\
  \bibinfo {pages} {090501} (\bibinfo {year} {2021})}\BibitemShut {NoStop}%
\bibitem [{\citenamefont {Ling}\ and\ \citenamefont {Kain}(2022)}]{Ling2022}%
  \BibitemOpen
  \bibfield  {author} {\bibinfo {author} {\bibfnamefont {H.~Y.}\ \bibnamefont
  {Ling}}\ and\ \bibinfo {author} {\bibfnamefont {B.}~\bibnamefont {Kain}},\
  }\bibfield  {title} {\bibinfo {title} {Topological study of a {Bogoliubov}-de
  {Gennes} system of pseudo spin-$1/2$ bosons with conserved magnetization in a
  honeycomb lattice},\ }\href {https://doi.org/10.1103/PhysRevA.105.023319}
  {\bibfield  {journal} {\bibinfo  {journal} {Phys. Rev. A}\ }\textbf {\bibinfo
  {volume} {105}},\ \bibinfo {pages} {023319} (\bibinfo {year}
  {2022})}\BibitemShut {NoStop}%
\bibitem [{\citenamefont {Hashimoto}\ \emph {et~al.}(2015)\citenamefont
  {Hashimoto}, \citenamefont {Kanki}, \citenamefont {Hayakawa},\ and\
  \citenamefont {Petrosky}}]{Hashimoto2015}%
  \BibitemOpen
  \bibfield  {author} {\bibinfo {author} {\bibfnamefont {K.}~\bibnamefont
  {Hashimoto}}, \bibinfo {author} {\bibfnamefont {K.}~\bibnamefont {Kanki}},
  \bibinfo {author} {\bibfnamefont {H.}~\bibnamefont {Hayakawa}},\ and\
  \bibinfo {author} {\bibfnamefont {T.}~\bibnamefont {Petrosky}},\ }\bibfield
  {title} {\bibinfo {title} {Non-divergent representation of a non-{Hermitian}
  operator near the exceptional point with application to a quantum {Lorentz}
  gas},\ }\href {https://doi.org/10.1093/ptep/ptu183} {\bibfield  {journal}
  {\bibinfo  {journal} {Prog. Theor. Exp. Phys.}\ }\textbf {\bibinfo {volume}
  {2015}},\ \bibinfo {pages} {23A02} (\bibinfo {year} {2015})}\BibitemShut
  {NoStop}%
\bibitem [{\citenamefont {Roccati}\ \emph {et~al.}(2022)\citenamefont
  {Roccati}, \citenamefont {Palma}, \citenamefont {Bagarello},\ and\
  \citenamefont {Ciccarello}}]{Roccati2022}%
  \BibitemOpen
  \bibfield  {author} {\bibinfo {author} {\bibfnamefont {F.}~\bibnamefont
  {Roccati}}, \bibinfo {author} {\bibfnamefont {G.~M.}\ \bibnamefont {Palma}},
  \bibinfo {author} {\bibfnamefont {F.}~\bibnamefont {Bagarello}},\ and\
  \bibinfo {author} {\bibfnamefont {F.}~\bibnamefont {Ciccarello}},\ }\bibfield
   {title} {\bibinfo {title} {Non-{Hermitian} physics and master equations},\
  }\Eprint {https://arxiv.org/abs/2201.05367} {arXiv:2201.05367}  (\bibinfo
  {year} {2022})\BibitemShut {NoStop}%
\bibitem [{\citenamefont {Yao}\ and\ \citenamefont {Wang}(2018)}]{Yao2018}%
  \BibitemOpen
  \bibfield  {author} {\bibinfo {author} {\bibfnamefont {S.}~\bibnamefont
  {Yao}}\ and\ \bibinfo {author} {\bibfnamefont {Z.}~\bibnamefont {Wang}},\
  }\bibfield  {title} {\bibinfo {title} {Edge states and topological invariants
  of non-{Hermitian} systems},\ }\href
  {https://doi.org/10.1103/PhysRevLett.121.086803} {\bibfield  {journal}
  {\bibinfo  {journal} {Phys. Rev. Lett.}\ }\textbf {\bibinfo {volume} {121}},\
  \bibinfo {pages} {086803} (\bibinfo {year} {2018})}\BibitemShut {NoStop}%
\bibitem [{\citenamefont {Li}\ \emph {et~al.}(2020)\citenamefont {Li},
  \citenamefont {Lee}, \citenamefont {Mu},\ and\ \citenamefont
  {Gong}}]{Li2020}%
  \BibitemOpen
  \bibfield  {author} {\bibinfo {author} {\bibfnamefont {L.}~\bibnamefont
  {Li}}, \bibinfo {author} {\bibfnamefont {C.~H.}\ \bibnamefont {Lee}},
  \bibinfo {author} {\bibfnamefont {S.}~\bibnamefont {Mu}},\ and\ \bibinfo
  {author} {\bibfnamefont {J.}~\bibnamefont {Gong}},\ }\bibfield  {title}
  {\bibinfo {title} {Critical non-{Hermitian} skin effect},\ }\href
  {https://doi.org/10.1038/s41467-020-18917-4} {\bibfield  {journal} {\bibinfo
  {journal} {Nat. Commun.}\ }\textbf {\bibinfo {volume} {11}},\ \bibinfo
  {pages} {1} (\bibinfo {year} {2020})}\BibitemShut {NoStop}%
\bibitem [{\citenamefont {Alsallom}\ \emph {et~al.}(2021)\citenamefont
  {Alsallom}, \citenamefont {Herviou}, \citenamefont {Yazyev},\ and\
  \citenamefont {Brzezi{\'n}ska}}]{Alsallom2021}%
  \BibitemOpen
  \bibfield  {author} {\bibinfo {author} {\bibfnamefont {F.}~\bibnamefont
  {Alsallom}}, \bibinfo {author} {\bibfnamefont {L.}~\bibnamefont {Herviou}},
  \bibinfo {author} {\bibfnamefont {O.~V.}\ \bibnamefont {Yazyev}},\ and\
  \bibinfo {author} {\bibfnamefont {M.}~\bibnamefont {Brzezi{\'n}ska}},\
  }\bibfield  {title} {\bibinfo {title} {Fate of the non-{Hermitian} skin
  effect in many-body fermionic systems},\ }\Eprint
  {https://arxiv.org/abs/2110.13164} {arXiv:2110.13164}  (\bibinfo {year}
  {2021})\BibitemShut {NoStop}%
\bibitem [{\citenamefont {Zhang}\ \emph
  {et~al.}(2020{\natexlab{b}})\citenamefont {Zhang}, \citenamefont {Yang},\
  and\ \citenamefont {Fang}}]{Zhang2020b}%
  \BibitemOpen
  \bibfield  {author} {\bibinfo {author} {\bibfnamefont {K.}~\bibnamefont
  {Zhang}}, \bibinfo {author} {\bibfnamefont {Z.}~\bibnamefont {Yang}},\ and\
  \bibinfo {author} {\bibfnamefont {C.}~\bibnamefont {Fang}},\ }\bibfield
  {title} {\bibinfo {title} {Correspondence between winding numbers and skin
  modes in non-{Hermitian} systems},\ }\href
  {https://doi.org/10.1103/PhysRevLett.125.126402} {\bibfield  {journal}
  {\bibinfo  {journal} {Phys. Rev. Lett.}\ }\textbf {\bibinfo {volume} {125}},\
  \bibinfo {pages} {126402} (\bibinfo {year} {2020}{\natexlab{b}})}\BibitemShut
  {NoStop}%
\bibitem [{\citenamefont {Song}\ \emph {et~al.}(2019)\citenamefont {Song},
  \citenamefont {Yao},\ and\ \citenamefont {Wang}}]{Song2019}%
  \BibitemOpen
  \bibfield  {author} {\bibinfo {author} {\bibfnamefont {F.}~\bibnamefont
  {Song}}, \bibinfo {author} {\bibfnamefont {S.}~\bibnamefont {Yao}},\ and\
  \bibinfo {author} {\bibfnamefont {Z.}~\bibnamefont {Wang}},\ }\bibfield
  {title} {\bibinfo {title} {Non-{Hermitian} skin effect and chiral damping in
  open quantum systems},\ }\href
  {https://doi.org/10.1103/PhysRevLett.123.170401} {\bibfield  {journal}
  {\bibinfo  {journal} {Phys. Rev. Lett.}\ }\textbf {\bibinfo {volume} {123}},\
  \bibinfo {pages} {170401} (\bibinfo {year} {2019})}\BibitemShut {NoStop}%
\bibitem [{\citenamefont {Lee}\ \emph {et~al.}(2020)\citenamefont {Lee},
  \citenamefont {Lee},\ and\ \citenamefont {Yang}}]{Lee2020}%
  \BibitemOpen
  \bibfield  {author} {\bibinfo {author} {\bibfnamefont {E.}~\bibnamefont
  {Lee}}, \bibinfo {author} {\bibfnamefont {H.}~\bibnamefont {Lee}},\ and\
  \bibinfo {author} {\bibfnamefont {B.~J.}\ \bibnamefont {Yang}},\ }\bibfield
  {title} {\bibinfo {title} {Many-body approach to non-{Hermitian} physics in
  fermionic systems},\ }\href {https://doi.org/10.1103/PhysRevB.101.121109}
  {\bibfield  {journal} {\bibinfo  {journal} {Phys. Rev. B}\ }\textbf {\bibinfo
  {volume} {101}},\ \bibinfo {pages} {121109} (\bibinfo {year}
  {2020})}\BibitemShut {NoStop}%
\bibitem [{\citenamefont {Yi}\ and\ \citenamefont {Yang}(2020)}]{Yi2020}%
  \BibitemOpen
  \bibfield  {author} {\bibinfo {author} {\bibfnamefont {Y.}~\bibnamefont
  {Yi}}\ and\ \bibinfo {author} {\bibfnamefont {Z.}~\bibnamefont {Yang}},\
  }\bibfield  {title} {\bibinfo {title} {Non-{Hermitian} skin modes induced by
  on-site dissipations and chiral tunneling effect},\ }\href
  {https://doi.org/10.1103/PhysRevLett.125.186802} {\bibfield  {journal}
  {\bibinfo  {journal} {Phys. Rev. Lett.}\ }\textbf {\bibinfo {volume} {125}},\
  \bibinfo {pages} {186802} (\bibinfo {year} {2020})}\BibitemShut {NoStop}%
\bibitem [{\citenamefont {Herviou}\ \emph
  {et~al.}(2019{\natexlab{a}})\citenamefont {Herviou}, \citenamefont
  {Bardarson},\ and\ \citenamefont {Regnault}}]{Herviou2019a}%
  \BibitemOpen
  \bibfield  {author} {\bibinfo {author} {\bibfnamefont {L.}~\bibnamefont
  {Herviou}}, \bibinfo {author} {\bibfnamefont {J.~H.}\ \bibnamefont
  {Bardarson}},\ and\ \bibinfo {author} {\bibfnamefont {N.}~\bibnamefont
  {Regnault}},\ }\bibfield  {title} {\bibinfo {title} {Defining a bulk-edge
  correspondence for non-{Hermitian} {Hamiltonians} via singular-value
  decomposition},\ }\href {https://doi.org/10.1103/PhysRevA.99.052118}
  {\bibfield  {journal} {\bibinfo  {journal} {Phys. Rev. A}\ }\textbf {\bibinfo
  {volume} {99}},\ \bibinfo {pages} {052118} (\bibinfo {year}
  {2019}{\natexlab{a}})}\BibitemShut {NoStop}%
\bibitem [{\citenamefont {Chen}\ \emph {et~al.}(2022)\citenamefont {Chen},
  \citenamefont {Peng}, \citenamefont {Lu},\ and\ \citenamefont
  {Lu}}]{Chen2022}%
  \BibitemOpen
  \bibfield  {author} {\bibinfo {author} {\bibfnamefont {W.}~\bibnamefont
  {Chen}}, \bibinfo {author} {\bibfnamefont {L.}~\bibnamefont {Peng}}, \bibinfo
  {author} {\bibfnamefont {H.}~\bibnamefont {Lu}},\ and\ \bibinfo {author}
  {\bibfnamefont {X.}~\bibnamefont {Lu}},\ }\bibfield  {title} {\bibinfo
  {title} {Characterizing bulk-boundary correspondence of one-dimensional
  non-{Hermitian} interacting systems by edge entanglement entropy},\ }\href
  {https://doi.org/10.1103/PhysRevB.105.075126} {\bibfield  {journal} {\bibinfo
   {journal} {Phys. Rev. B}\ }\textbf {\bibinfo {volume} {105}},\ \bibinfo
  {pages} {075126} (\bibinfo {year} {2022})}\BibitemShut {NoStop}%
\bibitem [{\citenamefont {Chang}\ \emph {et~al.}(2020)\citenamefont {Chang},
  \citenamefont {You}, \citenamefont {Wen},\ and\ \citenamefont
  {Ryu}}]{Chang2020}%
  \BibitemOpen
  \bibfield  {author} {\bibinfo {author} {\bibfnamefont {P.~Y.}\ \bibnamefont
  {Chang}}, \bibinfo {author} {\bibfnamefont {J.~S.}\ \bibnamefont {You}},
  \bibinfo {author} {\bibfnamefont {X.}~\bibnamefont {Wen}},\ and\ \bibinfo
  {author} {\bibfnamefont {S.}~\bibnamefont {Ryu}},\ }\bibfield  {title}
  {\bibinfo {title} {Entanglement spectrum and entropy in topological
  non-{Hermitian} systems and nonunitary conformal field theory},\ }\href
  {https://doi.org/10.1103/physrevresearch.2.033069} {\bibfield  {journal}
  {\bibinfo  {journal} {Phys. Rev. Res.}\ }\textbf {\bibinfo {volume} {2}},\
  \bibinfo {pages} {1} (\bibinfo {year} {2020})}\BibitemShut {NoStop}%
\bibitem [{\citenamefont {Herviou}\ \emph
  {et~al.}(2019{\natexlab{b}})\citenamefont {Herviou}, \citenamefont
  {Regnault},\ and\ \citenamefont {Bardarson}}]{Herviou2019}%
  \BibitemOpen
  \bibfield  {author} {\bibinfo {author} {\bibfnamefont {L.}~\bibnamefont
  {Herviou}}, \bibinfo {author} {\bibfnamefont {N.}~\bibnamefont {Regnault}},\
  and\ \bibinfo {author} {\bibfnamefont {J.~H.}\ \bibnamefont {Bardarson}},\
  }\bibfield  {title} {\bibinfo {title} {Entanglement spectrum and symmetries
  in non-{Hermitian} fermionic non-interacting models},\ }\href
  {https://doi.org/10.21468/scipostphys.7.5.069} {\bibfield  {journal}
  {\bibinfo  {journal} {SciPost Phys.}\ }\textbf {\bibinfo {volume} {7}},\
  \bibinfo {pages} {69} (\bibinfo {year} {2019}{\natexlab{b}})}\BibitemShut
  {NoStop}%
\bibitem [{\citenamefont {Yokomizo}\ and\ \citenamefont
  {Murakami}(2019)}]{Yokomizo2019}%
  \BibitemOpen
  \bibfield  {author} {\bibinfo {author} {\bibfnamefont {K.}~\bibnamefont
  {Yokomizo}}\ and\ \bibinfo {author} {\bibfnamefont {S.}~\bibnamefont
  {Murakami}},\ }\bibfield  {title} {\bibinfo {title} {Non-{Bloch} band theory
  of non-{Hermitian} systems},\ }\href
  {https://doi.org/10.1103/PhysRevLett.123.066404} {\bibfield  {journal}
  {\bibinfo  {journal} {Phys. Rev. Lett.}\ }\textbf {\bibinfo {volume} {123}},\
  \bibinfo {pages} {066404} (\bibinfo {year} {2019})}\BibitemShut {NoStop}%
\bibitem [{\citenamefont {Yang}\ \emph {et~al.}(2020)\citenamefont {Yang},
  \citenamefont {Zhang}, \citenamefont {Fang},\ and\ \citenamefont
  {Hu}}]{Yang2020a}%
  \BibitemOpen
  \bibfield  {author} {\bibinfo {author} {\bibfnamefont {Z.}~\bibnamefont
  {Yang}}, \bibinfo {author} {\bibfnamefont {K.}~\bibnamefont {Zhang}},
  \bibinfo {author} {\bibfnamefont {C.}~\bibnamefont {Fang}},\ and\ \bibinfo
  {author} {\bibfnamefont {J.}~\bibnamefont {Hu}},\ }\bibfield  {title}
  {\bibinfo {title} {Non-{Hermitian} bulk-boundary correspondence and auxiliary
  generalized {Brillouin} zone theory},\ }\href
  {https://doi.org/10.1103/PhysRevLett.125.226402} {\bibfield  {journal}
  {\bibinfo  {journal} {Phys. Rev. Lett.}\ }\textbf {\bibinfo {volume} {125}},\
  \bibinfo {pages} {226402} (\bibinfo {year} {2020})}\BibitemShut {NoStop}%
\bibitem [{\citenamefont {Yokomizo}\ and\ \citenamefont
  {Murakami}(2020)}]{Yokomizo2020}%
  \BibitemOpen
  \bibfield  {author} {\bibinfo {author} {\bibfnamefont {K.}~\bibnamefont
  {Yokomizo}}\ and\ \bibinfo {author} {\bibfnamefont {S.}~\bibnamefont
  {Murakami}},\ }\bibfield  {title} {\bibinfo {title} {Non-{Bloch} band theory
  and bulk-edge correspondence in non-{Hermitian} systems},\ }\href
  {https://doi.org/10.1093/ptep/ptaa140} {\bibfield  {journal} {\bibinfo
  {journal} {Prog. Theor. Exp. Phys.}\ }\textbf {\bibinfo {volume} {2020}},\
  \bibinfo {pages} {12A102} (\bibinfo {year} {2020})}\BibitemShut {NoStop}%
\bibitem [{\citenamefont {Sun}\ \emph {et~al.}(2022)\citenamefont {Sun},
  \citenamefont {Tang},\ and\ \citenamefont {Kou}}]{Sun2022}%
  \BibitemOpen
  \bibfield  {author} {\bibinfo {author} {\bibfnamefont {G.}~\bibnamefont
  {Sun}}, \bibinfo {author} {\bibfnamefont {J.-C.}\ \bibnamefont {Tang}},\ and\
  \bibinfo {author} {\bibfnamefont {S.-P.}\ \bibnamefont {Kou}},\ }\bibfield
  {title} {\bibinfo {title} {Biorthogonal quantum criticality in
  non-{Hermitian} many-body systems},\ }\href
  {https://doi.org/10.1007/s11467-021-1126-1} {\bibfield  {journal} {\bibinfo
  {journal} {Front. Phys.}\ }\textbf {\bibinfo {volume} {17}},\ \bibinfo
  {pages} {33502} (\bibinfo {year} {2022})}\BibitemShut {NoStop}%
\bibitem [{\citenamefont {Hamazaki}\ \emph {et~al.}(2019)\citenamefont
  {Hamazaki}, \citenamefont {Kawabata},\ and\ \citenamefont
  {Ueda}}]{Hamazaki2019}%
  \BibitemOpen
  \bibfield  {author} {\bibinfo {author} {\bibfnamefont {R.}~\bibnamefont
  {Hamazaki}}, \bibinfo {author} {\bibfnamefont {K.}~\bibnamefont {Kawabata}},\
  and\ \bibinfo {author} {\bibfnamefont {M.}~\bibnamefont {Ueda}},\ }\bibfield
  {title} {\bibinfo {title} {Non-{Hermitian} many-body localization},\ }\href
  {https://doi.org/10.1103/PhysRevLett.123.090603} {\bibfield  {journal}
  {\bibinfo  {journal} {Phys. Rev. Lett.}\ }\textbf {\bibinfo {volume} {123}},\
  \bibinfo {pages} {090603} (\bibinfo {year} {2019})}\BibitemShut {NoStop}%
\bibitem [{\citenamefont {Verstraete}\ and\ \citenamefont
  {Cirac}(2006)}]{Verstraete2006}%
  \BibitemOpen
  \bibfield  {author} {\bibinfo {author} {\bibfnamefont {F.}~\bibnamefont
  {Verstraete}}\ and\ \bibinfo {author} {\bibfnamefont {J.~I.}\ \bibnamefont
  {Cirac}},\ }\bibfield  {title} {\bibinfo {title} {Matrix product states
  represent ground states faithfully},\ }\href
  {https://doi.org/10.1103/PhysRevB.73.094423} {\bibfield  {journal} {\bibinfo
  {journal} {Phys. Rev. B}\ }\textbf {\bibinfo {volume} {73}},\ \bibinfo
  {pages} {094423} (\bibinfo {year} {2006})}\BibitemShut {NoStop}%
\bibitem [{\citenamefont {Hastings}(2007)}]{Hastings2007}%
  \BibitemOpen
  \bibfield  {author} {\bibinfo {author} {\bibfnamefont {M.~B.}\ \bibnamefont
  {Hastings}},\ }\bibfield  {title} {\bibinfo {title} {An area law for
  one-dimensional quantum systems},\ }\href
  {https://doi.org/10.1088/1742-5468/2007/08/p08024} {\bibfield  {journal}
  {\bibinfo  {journal} {J. Stat. Mech.: Theory Exp.}\ }\textbf {\bibinfo
  {volume} {2007}}\bibinfo  {number} { (08)},\ \bibinfo {pages}
  {P08024}}\BibitemShut {NoStop}%
\bibitem [{\citenamefont {Verstraete}\ \emph {et~al.}(2004)\citenamefont
  {Verstraete}, \citenamefont {Porras},\ and\ \citenamefont
  {Cirac}}]{Verstraete2004}%
  \BibitemOpen
\bibfield  {number} {  }\bibfield  {author} {\bibinfo {author} {\bibfnamefont
  {F.}~\bibnamefont {Verstraete}}, \bibinfo {author} {\bibfnamefont
  {D.}~\bibnamefont {Porras}},\ and\ \bibinfo {author} {\bibfnamefont {J.~I.}\
  \bibnamefont {Cirac}},\ }\bibfield  {title} {\bibinfo {title} {Density matrix
  renormalization group and periodic boundary conditions: A quantum information
  perspective},\ }\href {https://doi.org/10.1103/PhysRevLett.93.227205}
  {\bibfield  {journal} {\bibinfo  {journal} {Phys. Rev. Lett.}\ }\textbf
  {\bibinfo {volume} {93}},\ \bibinfo {pages} {227205} (\bibinfo {year}
  {2004})}\BibitemShut {NoStop}%
\bibitem [{\citenamefont {Schollw{\"o}ck}(2011)}]{Schollwoeck2011}%
  \BibitemOpen
  \bibfield  {author} {\bibinfo {author} {\bibfnamefont {U.}~\bibnamefont
  {Schollw{\"o}ck}},\ }\bibfield  {title} {\bibinfo {title} {The density-matrix
  renormalization group in the age of matrix product states},\ }\href
  {https://doi.org/10.1016/j.aop.2010.09.012} {\bibfield  {journal} {\bibinfo
  {journal} {Ann. Phys. (N. Y.)}\ }\textbf {\bibinfo {volume} {326}},\ \bibinfo
  {pages} {96} (\bibinfo {year} {2011})}\BibitemShut {NoStop}%
\bibitem [{\citenamefont {White}(1992)}]{White1992}%
  \BibitemOpen
  \bibfield  {author} {\bibinfo {author} {\bibfnamefont {S.~R.}\ \bibnamefont
  {White}},\ }\bibfield  {title} {\bibinfo {title} {Density matrix formulation
  for quantum renormalization groups},\ }\href
  {https://doi.org/10.1103/PhysRevLett.69.2863} {\bibfield  {journal} {\bibinfo
   {journal} {Phys. Rev. Lett.}\ }\textbf {\bibinfo {volume} {69}},\ \bibinfo
  {pages} {2863} (\bibinfo {year} {1992})},\ \bibinfo {note} {zSCC:
  0007412}\BibitemShut {NoStop}%
\bibitem [{\citenamefont {Carlon}\ \emph {et~al.}(1999)\citenamefont {Carlon},
  \citenamefont {Henkel},\ and\ \citenamefont {Schollw{\"{o}}ck}}]{Carlon1999}%
  \BibitemOpen
  \bibfield  {author} {\bibinfo {author} {\bibfnamefont {E.}~\bibnamefont
  {Carlon}}, \bibinfo {author} {\bibfnamefont {M.}~\bibnamefont {Henkel}},\
  and\ \bibinfo {author} {\bibfnamefont {U.}~\bibnamefont {Schollw{\"{o}}ck}},\
  }\bibfield  {title} {\bibinfo {title} {Density matrix renormalization group
  and reaction-diffusion processes},\ }\href
  {https://doi.org/10.1007/s100510050983} {\bibfield  {journal} {\bibinfo
  {journal} {Eur. Phys. J. B}\ }\textbf {\bibinfo {volume} {12}},\ \bibinfo
  {pages} {99} (\bibinfo {year} {1999})}\BibitemShut {NoStop}%
\bibitem [{\citenamefont {Hieida}(1998)}]{Hieida1998}%
  \BibitemOpen
  \bibfield  {author} {\bibinfo {author} {\bibfnamefont {Y.}~\bibnamefont
  {Hieida}},\ }\bibfield  {title} {\bibinfo {title} {Application of the density
  matrix renormalization group method to a non-equilibrium problem},\ }\href
  {https://doi.org/10.1143/jpsj.67.369} {\bibfield  {journal} {\bibinfo
  {journal} {J. Phys. Soc. Jpn.}\ }\textbf {\bibinfo {volume} {67}},\ \bibinfo
  {pages} {369} (\bibinfo {year} {1998})}\BibitemShut {NoStop}%
\bibitem [{\citenamefont {Rotureau}\ \emph {et~al.}(2006)\citenamefont
  {Rotureau}, \citenamefont {Michel}, \citenamefont {Nazarewicz}, \citenamefont
  {P{\l}oszajczak},\ and\ \citenamefont {Dukelsky}}]{Rotureau2006}%
  \BibitemOpen
  \bibfield  {author} {\bibinfo {author} {\bibfnamefont {J.}~\bibnamefont
  {Rotureau}}, \bibinfo {author} {\bibfnamefont {N.}~\bibnamefont {Michel}},
  \bibinfo {author} {\bibfnamefont {W.}~\bibnamefont {Nazarewicz}}, \bibinfo
  {author} {\bibfnamefont {M.}~\bibnamefont {P{\l}oszajczak}},\ and\ \bibinfo
  {author} {\bibfnamefont {J.}~\bibnamefont {Dukelsky}},\ }\bibfield  {title}
  {\bibinfo {title} {Density matrix renormalization group approach for
  many-body open quantum systems},\ }\href
  {https://doi.org/10.1103/physrevlett.97.110603} {\bibfield  {journal}
  {\bibinfo  {journal} {Phys. Rev. Lett.}\ }\textbf {\bibinfo {volume} {97}},\
  \bibinfo {pages} {110603} (\bibinfo {year} {2006})}\BibitemShut {NoStop}%
\bibitem [{\citenamefont {Yamamoto}\ \emph {et~al.}(2022)\citenamefont
  {Yamamoto}, \citenamefont {Nakagawa}, \citenamefont {Tezuka}, \citenamefont
  {Ueda},\ and\ \citenamefont {Kawakami}}]{Yamamoto2022}%
  \BibitemOpen
  \bibfield  {author} {\bibinfo {author} {\bibfnamefont {K.}~\bibnamefont
  {Yamamoto}}, \bibinfo {author} {\bibfnamefont {M.}~\bibnamefont {Nakagawa}},
  \bibinfo {author} {\bibfnamefont {M.}~\bibnamefont {Tezuka}}, \bibinfo
  {author} {\bibfnamefont {M.}~\bibnamefont {Ueda}},\ and\ \bibinfo {author}
  {\bibfnamefont {N.}~\bibnamefont {Kawakami}},\ }\bibfield  {title} {\bibinfo
  {title} {Universal properties of dissipative {Tomonaga-Luttinger} liquids:
  Case study of a non-{Hermitian} {XXZ} spin chain},\ }\href
  {https://doi.org/10.1103/physrevb.105.205125} {\bibfield  {journal} {\bibinfo
   {journal} {Phys. Rev. B}\ }\textbf {\bibinfo {volume} {105}},\ \bibinfo
  {pages} {205125} (\bibinfo {year} {2022})}\BibitemShut {NoStop}%
\bibitem [{\citenamefont {Zhang}\ \emph
  {et~al.}(2020{\natexlab{c}})\citenamefont {Zhang}, \citenamefont {Chen},
  \citenamefont {Zhang}, \citenamefont {Lang}, \citenamefont {Li},\ and\
  \citenamefont {Zhu}}]{Zhang2020c}%
  \BibitemOpen
  \bibfield  {author} {\bibinfo {author} {\bibfnamefont {D.-W.}\ \bibnamefont
  {Zhang}}, \bibinfo {author} {\bibfnamefont {Y.-L.}\ \bibnamefont {Chen}},
  \bibinfo {author} {\bibfnamefont {G.-Q.}\ \bibnamefont {Zhang}}, \bibinfo
  {author} {\bibfnamefont {L.-J.}\ \bibnamefont {Lang}}, \bibinfo {author}
  {\bibfnamefont {Z.}~\bibnamefont {Li}},\ and\ \bibinfo {author}
  {\bibfnamefont {S.-L.}\ \bibnamefont {Zhu}},\ }\bibfield  {title} {\bibinfo
  {title} {Skin superfluid, topological {Mott} insulators, and asymmetric
  dynamics in an interacting non-{Hermitian} {Aubry-Andr{\'{e}}-Harper}
  model},\ }\href {https://doi.org/10.1103/physrevb.101.235150} {\bibfield
  {journal} {\bibinfo  {journal} {Phys. Rev. B}\ }\textbf {\bibinfo {volume}
  {101}},\ \bibinfo {pages} {235150} (\bibinfo {year}
  {2020}{\natexlab{c}})}\BibitemShut {NoStop}%
\bibitem [{\citenamefont {Wang}\ and\ \citenamefont {Xiang}(1997)}]{Wang1997}%
  \BibitemOpen
  \bibfield  {author} {\bibinfo {author} {\bibfnamefont {X.}~\bibnamefont
  {Wang}}\ and\ \bibinfo {author} {\bibfnamefont {T.}~\bibnamefont {Xiang}},\
  }\bibfield  {title} {\bibinfo {title} {Transfer-matrix density-matrix
  renormalization-group theory for thermodynamics of one-dimensional quantum
  systems},\ }\href {https://doi.org/10.1103/PhysRevB.56.5061} {\bibfield
  {journal} {\bibinfo  {journal} {Phys. Rev. B}\ }\textbf {\bibinfo {volume}
  {56}},\ \bibinfo {pages} {5061} (\bibinfo {year} {1997})},\ \bibinfo {note}
  {zSCC: 0000319}\BibitemShut {NoStop}%
\bibitem [{\citenamefont {Enss}\ and\ \citenamefont
  {Schollw{\"o}ck}(2001)}]{Enss2001}%
  \BibitemOpen
  \bibfield  {author} {\bibinfo {author} {\bibfnamefont {T.}~\bibnamefont
  {Enss}}\ and\ \bibinfo {author} {\bibfnamefont {U.}~\bibnamefont
  {Schollw{\"o}ck}},\ }\bibfield  {title} {\bibinfo {title} {On the choice of
  the density matrix in the stochastic {TMRG}},\ }\href
  {https://doi.org/10.1088/0305-4470/34/38/305} {\bibfield  {journal} {\bibinfo
   {journal} {J. Phys. A: Math. Gen.}\ }\textbf {\bibinfo {volume} {34}},\
  \bibinfo {pages} {7769} (\bibinfo {year} {2001})}\BibitemShut {NoStop}%
\bibitem [{\citenamefont {Huang}(2011{\natexlab{a}})}]{Huang2011a}%
  \BibitemOpen
  \bibfield  {author} {\bibinfo {author} {\bibfnamefont {Y.-K.}\ \bibnamefont
  {Huang}},\ }\bibfield  {title} {\bibinfo {title} {Biorthonormal
  transfer-matrix renormalization-group method for non-{Hermitian} matrices},\
  }\href {https://doi.org/10.1103/PhysRevE.83.036702} {\bibfield  {journal}
  {\bibinfo  {journal} {Phys. Rev. E}\ }\textbf {\bibinfo {volume} {83}},\
  \bibinfo {pages} {036702} (\bibinfo {year} {2011}{\natexlab{a}})}\BibitemShut
  {NoStop}%
\bibitem [{\citenamefont {Peschel}\ and\ \citenamefont {Kaulke}()}]{Peschel}%
  \BibitemOpen
  \bibfield  {author} {\bibinfo {author} {\bibfnamefont {I.}~\bibnamefont
  {Peschel}}\ and\ \bibinfo {author} {\bibfnamefont {M.}~\bibnamefont
  {Kaulke}},\ }\bibfield  {title} {\bibinfo {title} {Non-{Hermitian} problems
  and some other aspects},\ }in\ \href {https://doi.org/10.1007/bfb0106078}
  {\emph {\bibinfo {booktitle} {Density-Matrix Renormalization}}}\ (\bibinfo
  {publisher} {Springer Berlin Heidelberg})\ pp.\ \bibinfo {pages}
  {279--285}\BibitemShut {NoStop}%
\bibitem [{\citenamefont {Nishino}\ and\ \citenamefont
  {Shibata}(1999)}]{Nishino1999}%
  \BibitemOpen
  \bibfield  {author} {\bibinfo {author} {\bibfnamefont {T.}~\bibnamefont
  {Nishino}}\ and\ \bibinfo {author} {\bibfnamefont {N.}~\bibnamefont
  {Shibata}},\ }\bibfield  {title} {\bibinfo {title} {Efficiency of asymmetric
  targeting for finite-{T} {DMRG}},\ }\href
  {https://doi.org/10.1143/jpsj.68.3501} {\bibfield  {journal} {\bibinfo
  {journal} {J. Phys. Soc. Jpn.}\ }\textbf {\bibinfo {volume} {68}},\ \bibinfo
  {pages} {3501} (\bibinfo {year} {1999})}\BibitemShut {NoStop}%
\bibitem [{\citenamefont {Chan}\ and\ \citenamefont
  {Voorhis}(2005)}]{Chan2005}%
  \BibitemOpen
  \bibfield  {author} {\bibinfo {author} {\bibfnamefont {G.~K.-L.}\
  \bibnamefont {Chan}}\ and\ \bibinfo {author} {\bibfnamefont {T.~V.}\
  \bibnamefont {Voorhis}},\ }\bibfield  {title} {\bibinfo {title}
  {Density-matrix renormalization-group algorithms with nonorthogonal orbitals
  and non-{Hermitian} operators, and applications to polyenes},\ }\href
  {https://doi.org/10.1063/1.1899124} {\bibfield  {journal} {\bibinfo
  {journal} {J. Chem. Phys.}\ }\textbf {\bibinfo {volume} {122}},\ \bibinfo
  {pages} {204101} (\bibinfo {year} {2005})}\BibitemShut {NoStop}%
\bibitem [{\citenamefont {Huang}(2011{\natexlab{b}})}]{Huang2011}%
  \BibitemOpen
  \bibfield  {author} {\bibinfo {author} {\bibfnamefont {Y.-K.}\ \bibnamefont
  {Huang}},\ }\bibfield  {title} {\bibinfo {title} {Biorthonormal
  matrix-product-state analysis for the non-{Hermitian} transfer-matrix
  renormalization group in the thermodynamic limit},\ }\href
  {https://doi.org/10.1088/1742-5468/2011/07/P07003} {\bibfield  {journal}
  {\bibinfo  {journal} {J. Stat. Mech.: Theory Exp.}\ }\textbf {\bibinfo
  {volume} {2011}}\bibinfo  {number} { (7)},\ \bibinfo {pages}
  {P07003}}\BibitemShut {NoStop}%
\bibitem [{sm()}]{sm}%
  \BibitemOpen
\bibfield  {number} {  }\href@noop {} {}\bibinfo {note} {See Supplemental
  Material for details.}\BibitemShut {Stop}%
\bibitem [{\citenamefont {Utreras-Alarc{\'{o}}n}\ \emph
  {et~al.}(2019)\citenamefont {Utreras-Alarc{\'{o}}n}, \citenamefont
  {Rivera-Tapia}, \citenamefont {Niklitschek},\ and\ \citenamefont
  {Delgado}}]{UtrerasAlarcon2019}%
  \BibitemOpen
  \bibfield  {author} {\bibinfo {author} {\bibfnamefont {A.}~\bibnamefont
  {Utreras-Alarc{\'{o}}n}}, \bibinfo {author} {\bibfnamefont {M.}~\bibnamefont
  {Rivera-Tapia}}, \bibinfo {author} {\bibfnamefont {S.}~\bibnamefont
  {Niklitschek}},\ and\ \bibinfo {author} {\bibfnamefont {A.}~\bibnamefont
  {Delgado}},\ }\bibfield  {title} {\bibinfo {title} {Stochastic optimization
  on complex variables and pure-state quantum tomography},\ }\href
  {https://doi.org/10.1038/s41598-019-52289-0} {\bibfield  {journal} {\bibinfo
  {journal} {Sci. Rep.}\ }\textbf {\bibinfo {volume} {9}},\ \bibinfo {pages}
  {1} (\bibinfo {year} {2019})}\BibitemShut {NoStop}%
\bibitem [{\citenamefont {Wirtinger}(1927)}]{Wirtinger1927}%
  \BibitemOpen
  \bibfield  {author} {\bibinfo {author} {\bibfnamefont {W.}~\bibnamefont
  {Wirtinger}},\ }\bibfield  {title} {\bibinfo {title} {Zur formalen theorie
  der funktionen von mehr komplexen {Ver{\"a}nderlichen}},\ }\href
  {https://doi.org/10.1007/BF01447872} {\bibfield  {journal} {\bibinfo
  {journal} {Math. Ann.}\ }\textbf {\bibinfo {volume} {97}},\ \bibinfo {pages}
  {357} (\bibinfo {year} {1927})}\BibitemShut {NoStop}%
\bibitem [{\citenamefont {Lieu}(2018)}]{Lieu2018}%
  \BibitemOpen
  \bibfield  {author} {\bibinfo {author} {\bibfnamefont {S.}~\bibnamefont
  {Lieu}},\ }\bibfield  {title} {\bibinfo {title} {Topological phases in the
  non-{Hermitian} {Su}-{Schrieffer}-{Heeger} model},\ }\href
  {https://doi.org/10.1103/PhysRevB.97.045106} {\bibfield  {journal} {\bibinfo
  {journal} {Phys. Rev. B}\ }\textbf {\bibinfo {volume} {97}},\ \bibinfo
  {pages} {045106} (\bibinfo {year} {2018})}\BibitemShut {NoStop}%
\bibitem [{\citenamefont {Han}\ \emph {et~al.}(2021)\citenamefont {Han},
  \citenamefont {Liu},\ and\ \citenamefont {Liu}}]{Han2021}%
  \BibitemOpen
  \bibfield  {author} {\bibinfo {author} {\bibfnamefont {Y.~Z.}\ \bibnamefont
  {Han}}, \bibinfo {author} {\bibfnamefont {J.~S.}\ \bibnamefont {Liu}},\ and\
  \bibinfo {author} {\bibfnamefont {C.~S.}\ \bibnamefont {Liu}},\ }\bibfield
  {title} {\bibinfo {title} {The topological counterparts of non-{Hermitian}
  {SSH} models},\ }\href {https://doi.org/10.1088/1367-2630/ac3e9f} {\bibfield
  {journal} {\bibinfo  {journal} {New J. Phys.}\ }\textbf {\bibinfo {volume}
  {23}},\ \bibinfo {pages} {123029} (\bibinfo {year} {2021})}\BibitemShut
  {NoStop}%
\bibitem [{\citenamefont {Xi}\ \emph {et~al.}(2021)\citenamefont {Xi},
  \citenamefont {Zhang}, \citenamefont {Gu},\ and\ \citenamefont
  {Chen}}]{Xi2021}%
  \BibitemOpen
  \bibfield  {author} {\bibinfo {author} {\bibfnamefont {W.}~\bibnamefont
  {Xi}}, \bibinfo {author} {\bibfnamefont {Z.-H.}\ \bibnamefont {Zhang}},
  \bibinfo {author} {\bibfnamefont {Z.-C.}\ \bibnamefont {Gu}},\ and\ \bibinfo
  {author} {\bibfnamefont {W.-Q.}\ \bibnamefont {Chen}},\ }\bibfield  {title}
  {\bibinfo {title} {Classification of topological phases in one dimensional
  interacting non-{Hermitian} systems and emergent unitarity},\ }\href
  {https://doi.org/10.1016/j.scib.2021.04.027} {\bibfield  {journal} {\bibinfo
  {journal} {Sci. Bull.}\ }\textbf {\bibinfo {volume} {66}},\ \bibinfo {pages}
  {1731} (\bibinfo {year} {2021})}\BibitemShut {NoStop}%
\bibitem [{\citenamefont {Fern{\'a}ndez}(2018)}]{Fernandez2018}%
  \BibitemOpen
  \bibfield  {author} {\bibinfo {author} {\bibfnamefont {F.~M.}\ \bibnamefont
  {Fern{\'a}ndez}},\ }\bibfield  {title} {\bibinfo {title} {Exceptional point
  in a simple textbook example},\ }\href
  {https://doi.org/10.1088/1361-6404/aab6df} {\bibfield  {journal} {\bibinfo
  {journal} {Eur. J. Phys.}\ }\textbf {\bibinfo {volume} {39}},\ \bibinfo
  {pages} {045005} (\bibinfo {year} {2018})}\BibitemShut {NoStop}%
\bibitem [{\citenamefont {Heiss}(2012)}]{Heiss2012}%
  \BibitemOpen
  \bibfield  {author} {\bibinfo {author} {\bibfnamefont {W.~D.}\ \bibnamefont
  {Heiss}},\ }\bibfield  {title} {\bibinfo {title} {The physics of exceptional
  points},\ }\href {https://doi.org/10.1088/1751-8113/45/44/444016} {\bibfield
  {journal} {\bibinfo  {journal} {J. Phys. A Math. Theor.}\ }\textbf {\bibinfo
  {volume} {45}},\ \bibinfo {pages} {444016} (\bibinfo {year}
  {2012})}\BibitemShut {NoStop}%
\bibitem [{\citenamefont {Tzeng}\ \emph {et~al.}(2021)\citenamefont {Tzeng},
  \citenamefont {Ju}, \citenamefont {Chen},\ and\ \citenamefont
  {Huang}}]{Tzeng2021}%
  \BibitemOpen
  \bibfield  {author} {\bibinfo {author} {\bibfnamefont {Y.~C.}\ \bibnamefont
  {Tzeng}}, \bibinfo {author} {\bibfnamefont {C.~Y.}\ \bibnamefont {Ju}},
  \bibinfo {author} {\bibfnamefont {G.~Y.}\ \bibnamefont {Chen}},\ and\
  \bibinfo {author} {\bibfnamefont {W.~M.}\ \bibnamefont {Huang}},\ }\bibfield
  {title} {\bibinfo {title} {Hunting for the non-{Hermitian} exceptional points
  with fidelity susceptibility},\ }\href
  {https://doi.org/10.1103/PhysRevResearch.3.013015} {\bibfield  {journal}
  {\bibinfo  {journal} {Phys. Rev. Res.}\ }\textbf {\bibinfo {volume} {3}},\
  \bibinfo {pages} {013015} (\bibinfo {year} {2021})}\BibitemShut {NoStop}%
\bibitem [{\citenamefont {Guo}\ \emph {et~al.}(2021)\citenamefont {Guo},
  \citenamefont {Yu}, \citenamefont {Huang}, \citenamefont {Yang},
  \citenamefont {Chi}, \citenamefont {Liao},\ and\ \citenamefont
  {Xiang}}]{Guo2021}%
  \BibitemOpen
  \bibfield  {author} {\bibinfo {author} {\bibfnamefont {Y.-B.}\ \bibnamefont
  {Guo}}, \bibinfo {author} {\bibfnamefont {Y.-C.}\ \bibnamefont {Yu}},
  \bibinfo {author} {\bibfnamefont {R.-Z.}\ \bibnamefont {Huang}}, \bibinfo
  {author} {\bibfnamefont {L.-P.}\ \bibnamefont {Yang}}, \bibinfo {author}
  {\bibfnamefont {R.-Z.}\ \bibnamefont {Chi}}, \bibinfo {author} {\bibfnamefont
  {H.-J.}\ \bibnamefont {Liao}},\ and\ \bibinfo {author} {\bibfnamefont
  {T.}~\bibnamefont {Xiang}},\ }\bibfield  {title} {\bibinfo {title}
  {Entanglement entropy of non-{Hermitian} free fermions},\ }\href
  {https://doi.org/10.1088/1361-648x/ac216e} {\bibfield  {journal} {\bibinfo
  {journal} {J. Phys. Condens. Matter}\ }\textbf {\bibinfo {volume} {33}},\
  \bibinfo {pages} {475502} (\bibinfo {year} {2021})}\BibitemShut {NoStop}%
\bibitem [{\citenamefont {Tu}\ \emph {et~al.}(2021)\citenamefont {Tu},
  \citenamefont {Tzeng},\ and\ \citenamefont {Chang}}]{Tu2021}%
  \BibitemOpen
  \bibfield  {author} {\bibinfo {author} {\bibfnamefont {Y.-T.}\ \bibnamefont
  {Tu}}, \bibinfo {author} {\bibfnamefont {Y.-C.}\ \bibnamefont {Tzeng}},\ and\
  \bibinfo {author} {\bibfnamefont {P.-Y.}\ \bibnamefont {Chang}},\ }\bibfield
  {title} {\bibinfo {title} {{R{\'e}nyi} entropies and negative central charges
  in non-{Hermitian} quantum systems},\ }\href
  {https://doi.org/10.21468/SciPostPhys.12.6.194} {\bibfield  {journal}
  {\bibinfo  {journal} {SciPost Phys.}\ }\textbf {\bibinfo {volume} {12}},\
  \bibinfo {pages} {194} (\bibinfo {year} {2021})}\BibitemShut {NoStop}%
\bibitem [{\citenamefont {Brody}(2013)}]{Brody2013}%
  \BibitemOpen
  \bibfield  {author} {\bibinfo {author} {\bibfnamefont {D.~C.}\ \bibnamefont
  {Brody}},\ }\bibfield  {title} {\bibinfo {title} {Biorthogonal quantum
  mechanics},\ }\href {https://doi.org/10.1088/1751-8113/47/3/035305}
  {\bibfield  {journal} {\bibinfo  {journal} {J. Phys. A Math. Theor.}\
  }\textbf {\bibinfo {volume} {47}},\ \bibinfo {pages} {035305} (\bibinfo
  {year} {2013})}\BibitemShut {NoStop}%
\bibitem [{\citenamefont {Pati}(2009)}]{Pati2009}%
  \BibitemOpen
  \bibfield  {author} {\bibinfo {author} {\bibfnamefont {A.~K.}\ \bibnamefont
  {Pati}},\ }\bibfield  {title} {\bibinfo {title} {Entanglement in
  non-{Hermitian} quantum theory},\ }\href
  {https://doi.org/10.1007/s12043-009-0101-0} {\bibfield  {journal} {\bibinfo
  {journal} {Pramana}\ }\textbf {\bibinfo {volume} {73}},\ \bibinfo {pages}
  {485} (\bibinfo {year} {2009})}\BibitemShut {NoStop}%
\bibitem [{\citenamefont {Korff}(2008)}]{Korff2008}%
  \BibitemOpen
  \bibfield  {author} {\bibinfo {author} {\bibfnamefont {C.}~\bibnamefont
  {Korff}},\ }\bibfield  {title} {\bibinfo {title} {{PT} symmetry of the
  non-{Hermitian} {XX} spin-chain: non-local bulk interaction from complex
  boundary fields},\ }\href {https://doi.org/10.1088/1751-8113/41/29/295206}
  {\bibfield  {journal} {\bibinfo  {journal} {J. Phys. A Math. Theor.}\
  }\textbf {\bibinfo {volume} {41}},\ \bibinfo {pages} {295206} (\bibinfo
  {year} {2008})}\BibitemShut {NoStop}%
\bibitem [{\citenamefont {Okuma}\ and\ \citenamefont {Sato}(2020)}]{Okuma2020}%
  \BibitemOpen
  \bibfield  {author} {\bibinfo {author} {\bibfnamefont {N.}~\bibnamefont
  {Okuma}}\ and\ \bibinfo {author} {\bibfnamefont {M.}~\bibnamefont {Sato}},\
  }\bibfield  {title} {\bibinfo {title} {{Hermitian} zero modes protected by
  nonnormality: Application of pseudospectra},\ }\href
  {https://doi.org/10.1103/PhysRevB.102.014203} {\bibfield  {journal} {\bibinfo
   {journal} {Phys. Rev. B}\ }\textbf {\bibinfo {volume} {102}},\ \bibinfo
  {pages} {014203} (\bibinfo {year} {2020})}\BibitemShut {NoStop}%
\bibitem [{\citenamefont {Rakov}(2018)}]{Rakov2018}%
  \BibitemOpen
  \bibfield  {author} {\bibinfo {author} {\bibfnamefont {M.~V.}\ \bibnamefont
  {Rakov}},\ }\bibfield  {title} {\bibinfo {title} {First-principle
  construction of {U}(1) symmetric matrix product states},\ }\href
  {https://doi.org/10.1007/s10909-018-1894-3} {\bibfield  {journal} {\bibinfo
  {journal} {J. Low Temp. Phys.}\ }\textbf {\bibinfo {volume} {192}},\ \bibinfo
  {pages} {75} (\bibinfo {year} {2018})}\BibitemShut {NoStop}%
\bibitem [{\citenamefont {Goldberg}(1991)}]{Goldberg1991}%
  \BibitemOpen
  \bibfield  {author} {\bibinfo {author} {\bibfnamefont {D.}~\bibnamefont
  {Goldberg}},\ }\bibfield  {title} {\bibinfo {title} {What every computer
  scientist should know about floating-point arithmetic},\ }\href
  {https://doi.org/10.1145/103162.103163} {\bibfield  {journal} {\bibinfo
  {journal} {ACM Comput. Surv.}\ }\textbf {\bibinfo {volume} {23}},\ \bibinfo
  {pages} {5} (\bibinfo {year} {1991})}\BibitemShut {NoStop}%
\end{thebibliography}%

\appendix

\section*{Supplemental material}

\section{Wirtinger derivative of energy}

The expectation energy of a Hermitian matrix $M$ with respect to a complex state $|x\rangle \triangleq |r\rangle+i|i\rangle$ is given by
\begin{equation}
\begin{aligned}
\mathcal{E}(|x\rangle) =& \frac{\langle x|M|x\rangle}{\langle x|x\rangle} \\
=& \frac{\langle r|M|r\rangle+\langle i|M|i\rangle+i\langle r|M|i\rangle - i\langle i|M|r\rangle}{\langle r|r\rangle + \langle i|i\rangle}.
\end{aligned}
\end{equation} 
Based on matrix calculus, we obtain
\begin{equation}
\begin{aligned}
&\partial_{\langle r|}\langle x|M|x\rangle = M^T|r\rangle+M|r\rangle+iM|i\rangle-iM^T|i\rangle,\\
&\partial_{\langle i|}\langle x|M|x\rangle = M^T|i\rangle+M|i\rangle+iM^T|r\rangle-iM|r\rangle,
\end{aligned}
\end{equation} and
\begin{equation}
\partial_{\langle r|}\langle x|x\rangle = 2|r\rangle,\;
\partial_{\langle i|}\langle x|x\rangle = 2|i\rangle.
\end{equation}
According to the definition of Wirtinger derivative, we have
\begin{equation}
\begin{aligned}
\partial_{\langle x|}\langle x|M|x\rangle&\triangleq\frac12\partial_{\langle r|}\langle x|M|x\rangle+\frac{i}2\partial_{\langle i|}\langle x|M|x\rangle =M|x\rangle\label{equ:xmx},
\end{aligned}
\end{equation}
\begin{equation}
\begin{aligned}
\partial_{\langle x|}\langle x|x\rangle&\triangleq\frac12\partial_{\langle r|}\langle x|x\rangle+\frac{i}2\partial_{\langle i|}\langle x|x\rangle =|x\rangle\label{equ:xx}.
\end{aligned}
\end{equation}
Combined with the chain rule it gives
\begin{equation}
\begin{aligned}
\partial_{\langle x|} \mathcal{E}(|x\rangle) =\frac{M|x\rangle}{\langle x|x\rangle} - \frac{\langle x|M|x\rangle|x\rangle}{\langle x|x\rangle^2}.\label{equ:wd2}
\end{aligned}
\end{equation}
The expectation energy of a non-Hermitian Hamiltonian $H$ is given by $e(|x \rangle) = \langle x|H|x\rangle/\langle x|x\rangle$.
By substituting $M=(H+H^\dag)/2$ into Eq.~(\ref{equ:wd2}), we obtain the derivative of the real part.
\begin{equation}
\begin{aligned}
\partial_{\langle x|}\Re\{e(|x\rangle)\}
&=\partial_{\langle x|}\frac{\langle x|(H+H^\dag)|x\rangle}{2\langle x|x\rangle} \\
&= \frac{(H+H^\dag )|x\rangle}{2\langle x|x\rangle} -\frac{\langle x|(H+H^\dag)|x\rangle|x\rangle}{2\langle x|x\rangle^2}.
\end{aligned}
\end{equation}
Similarly, the derivative of the imaginary part is obtained by replacing $M=(H-H^\dag)/(2i)$
\begin{equation}
    \begin{aligned}
        \partial_{\langle x|}\Im\{e(|x\rangle)\}&=\partial_{\langle x|}\frac{\langle x|(H-H^\dag)|x\rangle}{2i\langle x|x\rangle}\\&= \frac{(H-H^\dag )|x\rangle}{2i\langle x|x\rangle} -\frac{\langle x|(H-H^\dag)|x\rangle|x\rangle}{2i\langle x|x\rangle^2}.
    \end{aligned}
\end{equation}

\section{Wirtinger derivative of the eigenvector residual norm}

Given a Hermitian matrix $M$, the Wirtinger derivative of $\langle x|M|x\rangle^2$ is
\begin{equation}
\partial_{\langle x|}\langle x|M|x\rangle^2=2\langle x|M|x\rangle\partial_{\langle x|}\langle x|M|x\rangle=2\langle x|M|x\rangle M|x\rangle.
\end{equation}
Together with Eqs.~(\ref{equ:xmx}) and (\ref{equ:xx}) we obtain
\begin{equation}
\partial_{\langle x|}\frac{\langle x|M|x\rangle^2}{\langle x|x\rangle}=\frac{2\langle x|M|x\rangle M|x\rangle}{\langle x|x\rangle}- \frac{\langle x|M|x\rangle^2|x\rangle}{\langle x|x\rangle^2}.\label{equ:xmx2}
\end{equation}

The eigenvector residual norm is defined as \begin{equation}
\begin{aligned}
& \mathcal{N}(|x\rangle) \triangleq \left|H|x\rangle-\frac{\langle x|H|x\rangle}{\langle x|x\rangle}|x\rangle\right|^2\\
=& \langle x|H^\dag H|x\rangle-\frac{\langle x|H|x\rangle\langle x|H^\dag|x\rangle}{\langle x|x\rangle}\\
=& \langle x|H^\dag H|x\rangle-\frac{\langle x|(H+H^\dag)|x\rangle^2+\langle x|i(H-H^\dag)|x\rangle^2}{4\langle x|x\rangle}.
\end{aligned}
\end{equation}
In the last step above we have replaced the multiplication of a complex number $\langle x|H|x\rangle$ and its complex conjunction by the square summation of its real and imaginary parts. 
Since $H^\dag H$, $(H+H^\dag)$, and $i(H-H^\dag)$ are always Hermitian, using Eqs. (\ref{equ:xmx}) and (\ref{equ:xmx2}) we obtain the Wirtinger derivative of $\mathcal{N}(|x\rangle)$:
\begin{equation}
\begin{aligned}
& \partial_{\langle x|}\mathcal{N}(|x\rangle) \\
= & H^\dag H|x\rangle - \frac{\langle x|(H+H^\dag)|x\rangle(H+H^\dag)|x\rangle}{2\langle x|x\rangle}\\
&+ \frac{\langle x|(H-H^\dag)|x\rangle(H-H^\dag)|x\rangle}{2\langle x|x\rangle}\\
&+ \frac{\langle x|H|x\rangle\langle x|H^\dag|x\rangle|x\rangle}{\langle x|x\rangle^2}\\
= &\left(H^\dag -\frac{\langle x|H^\dag|x\rangle}{\langle x|x\rangle}\right)\left(H -\frac{\langle x|H|x\rangle}{\langle x|x\rangle}\right)|x\rangle.
\end{aligned}
\end{equation}

\section{Numerical instabilities of the bi-orthogonal condition}
\label{sec:bi}

\begin{figure}[tbp]
\centering
\includegraphics[width=1.0\columnwidth]{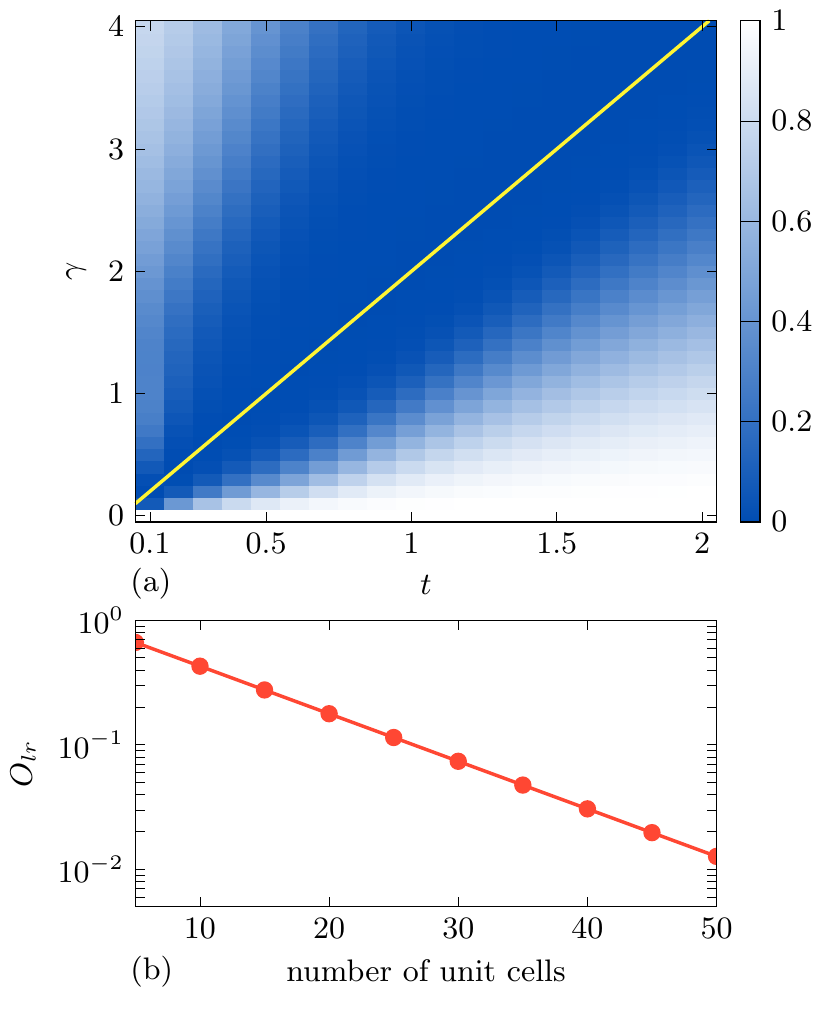}
\caption{
The bi-orthogonal overlap $O_{lr}$ between normalized left and right eigenstates for the eigenvalue with the smallest real part. 
(a) Results from exact diagonalization for the $6$-unit-cell nH-SSH model with open boundary condition. 
The yellow solid line denotes exceptional points. 
(b) Scaling of $O_{lr}$ with system size for $t=1.8$ and $\gamma=1.3$ using the HVMPS algorithm at virtual bond dimension $D=200$.
}
\label{fig:biorth}
\end{figure}

The bi-orthogonal formalism is an interesting approach for treating non-Hermitian systems. 
However, when the bi-orthogonal overlap between ground states $O_{lr}\triangleq |\langle l|r\rangle|/\sqrt{\langle l|l\rangle\langle r|r\rangle}$ approaches zero, the bi-orthogonal condition induces numerical instabilities, which is especially troublesome for large-system numerical studies.

We compute $O_{lr}$ for the $6$-unit-cell nH-SSH model using exact diagonalization, and the results are shown in Fig.~\ref{fig:biorth}(a). 
When approaching the exceptional points denoted by the yellow solid line, we clearly see that $O_{lr}$ steadily decreases to zero.
If the biorthogonal condition still requires $O_{lr}=1$, extreme large numbers will appear in $|l \rangle$ and $|r \rangle$, resulting in numerical instabilities.
When our HVMPS algorithm is used instead, the scaling behavior for larger systems is free of the aforementioned numerical difficulties.
As illustrated in Fig.~\ref{fig:biorth}(b), $O_{lr}$ decays exponentially while remaining accurate with growing system size.

The fast decay of $O_{lr}$ is an universal phenomenon in parity-time ($\mathcal{PT}$) symmetric system with a non-Hermitian Hamiltonian satisfying $PHP=H^\dag$, where $P$ denotes the spatial parity operator.
Combined with the eigen-decomposition
\begin{equation}
H=\sum\limits_i e_i|r_i \rangle\langle l_i |,
\end{equation} 
one finds that
\begin{equation}
H = {(PHP)}^\dag =\sum\limits_i e_i^*P|l_i \rangle\langle r_i|P,
\end{equation} which shows
\begin{equation}
\langle l_i| = \langle r_i|P\label{eq:p}
\end{equation}
for all modes with real eigenvalues. 
If the wave functions are localized at one end of the chain, $|\langle l_i|r_i \rangle| = |\langle r_i |P|r_i \rangle|$ will be small. 
This is thought to be a common feature of non-Hermitian systems and is called skin effect~\cite{Yao2018}, or hereby referred to a more general many-body skin effect.
It is worth noting that Eq.~(\ref{eq:p}) is valid for the many-body ground state $|\rm SR \rangle$ of a general $\mathcal{PT}$ symmetric Hamiltonian, which processes a real eigenvalue in both $\mathcal{PT}$ symmetry breaking and unbroken phases.

\section{The gradient variational matrix product state (GVMPS) algorithm}

\subsection{Starting point and the converged state}

For a general non-Hermitian Hamiltonian, the gradient descent generates a flow in the complex plane of energies, with eigenvalues of $H$ acting as attractors.
Starting from any position, the energy ends up at one of the attractors, and the converged eigenstate depends on the starting point.

In the main text, we demonstrate the ground enengy of $(H+H^\dag)/2$ is a good starting point to find $|\rm SR\rangle$ with the smallest real eigenvalue.
Likewise, the ground state energy of $(H-H^\dag)/2$ is an ideal beginning for determining $|\rm SI\rangle$ with the smallest imaginary eigenvalue.
The $|\rm SI\rangle$ ground state of $H$ is also the $|\rm SR\rangle$ ground state of $-iH$. 
Furthermore, we can find the eigenenergy with the largest absolute value in any direction in the complex plane by applying a global rotation to the whole spectra. 
Namely, solving the $|\rm SR\rangle$ ground state of $e^{i(\pi-\theta)}H$ yields the greatest absolute eigenenergy of $H$ in direction $\theta$.

\subsection{Gradient of the objective function $s_n(\varepsilon)$}
 
The objective function for gradient descent is chosen as the smallest singular value $s_n(\varepsilon)$ in singular value decomposition
\begin{equation}
H-\varepsilon=USV^\dag,
\label{equ:svd2}
\end{equation} 
where $U^\dag U=V^\dag V=I$. 
The differential form of the unitary condition is
\begin{equation}
\dif U^\dag U+U^\dag \dif U=\dif V^\dag V+V^\dag \dif V=0.
\end{equation} 
Therefore, $\dif U^\dag U$ and $\dif V^\dag V$ are skew-Hermitian with their diagonal elements pure imaginary. 
Left-multiplying by $U^\dag$ and right-multiplying by $V$ to the differential form of Eq.~(\ref{equ:svd2}) with respect to $\varepsilon$ gives
\begin{equation}
-U^\dag Vd\varepsilon=U^\dag \dif US+\dif S+S\dif V^\dag V.
\end{equation} 
By comparing the real parts of the $n$-th diagonal elements from both sides, $\dif s_n$ is found to satisfy
\begin{equation}
\dif s_n = -\Re\{U^\dag_n V_n \dif \varepsilon\}.
\label{dsn1}
\end{equation}
On the other hand, the differential of a real-valued function $s_n(\varepsilon)$ with a complex-valued argument $\varepsilon$ can be expressed as
\begin{equation}
\dif s_n =2 \Re\{\partial_{\varepsilon^*} s_n \dif \varepsilon^* \} = 2\Re\{(\partial_{\varepsilon^*}s_n)^* \dif \varepsilon\}.
\label{dsn2}
\end{equation}
Comparing Eq.~(\ref{dsn1}) with Eq.~(\ref{dsn2}) and using the relation $V^{\dagger} H V - \varepsilon = V^{\dagger} U S$, we obtain
\begin{equation}
\partial_{\varepsilon^*}s_n = -\frac{V^\dag_n U_n}2 = \frac{\varepsilon-V_n^\dag HV_n}{2s_n}.
\label{equ:dsde}
\end{equation}
where $V_n$ ($U_n$) is the $n$-th component of $V$ ($U$) with the smallest singular value $s_n$.
As $s_n$ approaches $0$, $V_n$ satisfies $(H-\varepsilon)V_n \approx 0$ while $U_n$ satisfies $U_n^{\dagger}(H-\varepsilon) \approx 0$, implying that $V_n$ and $U_n$ are approximately the right and left eigenvectors of $H$.
As mentioned in Appendix~\ref{sec:bi}, $U^\dag_n V_n$ decreases to zero exponentially with system size for the $\mathcal{PT}$ symmetric model, which slows down the gradient descent when approaching the minimum.

\subsection{A brief description of the VMPS method}

\begin{figure}[tbp]
\centering
\includegraphics[width=0.9\columnwidth]{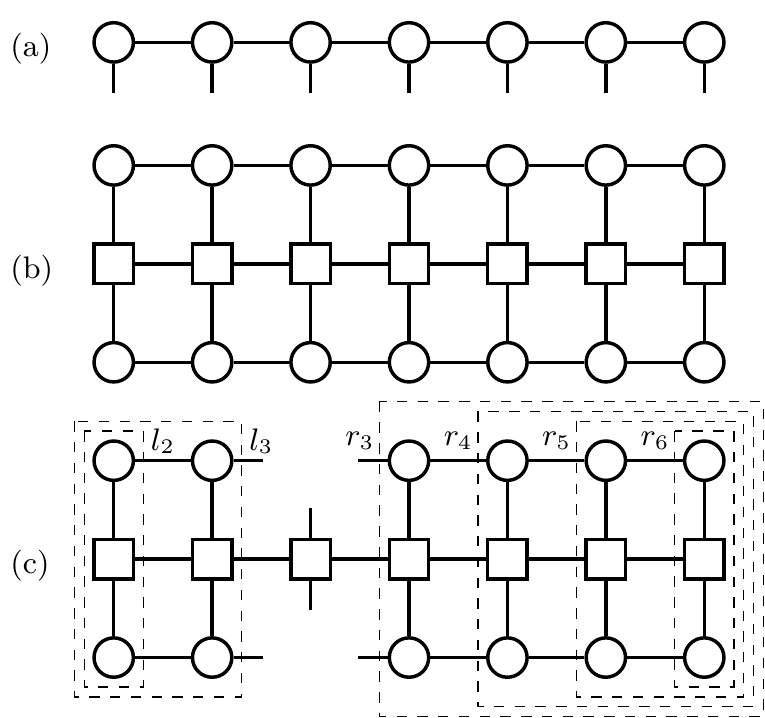}
\caption{
Graphic notations of (a) MPS, (b) energy, and (c) effective Hamiltonian $H_{\rm eff}$.
Here $l_i$ and $r_i$ are left and right environments.
}
\label{fig:tensor}
\end{figure}

Matrix product states can naturally represent the ground state of a $N$-site gapped system described by a local Hermitian Hamiltonian.
\begin{equation}
\begin{split}
|\psi\rangle=&\sum\limits_{\{p\}}\left[\sum\limits_{\{v\}}{(T_1)}^{p_1}_{v_1,v_2}{(T_2)}^{p_2}_{v_2,v_3}\cdots {(T_N)}^{p_N}_{v_N,v_{N+1}}\right]\\
& |p_1p_2\cdots p_N\rangle,
\end{split}
\end{equation}
with $\{p\}$ the physical indices and $\{v\}$ the virtual indices, as illustrated in Fig.~\ref{fig:tensor}(a).
A Hamiltonian can be represented by a matrix product operator (MPO)
\begin{equation}
\begin{aligned}
H = &\sum\limits_{\{p,q\}}\left[\sum\limits_{\{u\}}{(W_1)}^{q_1,p_1}_{u_1,u_2}{(W_2)}^{q_2,p_2}_{u_2,u_3}\cdots {(W_N)}^{q_N,p_N}_{u_N,u_{N+1}}\right]\\
& |q_1q_2\cdots q_N\rangle\langle p_1p_2\cdots p_N|,
\end{aligned}
\end{equation}
with $\{u\}$ the virtual indices and $\{p\} \{q\}$ the physical indices. 
For systems with OBC, only one virtual index is required for $v_1$, $v_{N+1}$, $u_1$, and $u_{N+1}$.
The energy $\langle \psi|H| \psi \rangle$ is expressed as Fig.~\ref{fig:tensor}(b) with the normalization constraint $\langle \psi | \psi \rangle=1$, where MPS with upward physical indices represents the state in bra space. 
Analogous to the standard variational method, VMPS minimizes energy site by site in the MPS ansatz. 
To optimize the $i$-th site, the effective Hamiltonian $H_{\rm eff}$ in Fig.~\ref{fig:tensor}(c) is obtained by contracting the left/right environment $l_i/r_i$ with the MPO $W_i$ at this site.
$H_{\rm eff}$ is viewed as a matrix with three downward indices forming a row index and three upward indices forming a column index. 
The ground eigenvector of $H_{\rm eff}$ that minimizes the energy in the current step is calculated and reshaped into a tensor $T_i'$ to replace the previous $T_i$ at this site. 
After that, the calculation continues to the next site, going forward and backward until convergence is reached. 

\subsection{GVMPS algorithm}

To change the Hamiltonian from $H$ to $(H-\varepsilon)$ in the MPO representation, one may simply change the MPO on an arbitrary single site $i$ from $W_i$ to $W_i-\varepsilon$ while leaving all others unchanged.
Thus, $(H^\dag-\varepsilon^*)(H-\varepsilon)$ and $H^\dag H$ only differ on one site.
To represent $\langle \psi|(H^\dag-\varepsilon^*)(H-\varepsilon) |\psi \rangle$, we may reuse the environment of $\langle \psi| H^{\dagger}H |\psi \rangle$ to reduce computational effort.
In the algorithm, the left/right environment for $\langle \psi| H^\dag H |\psi \rangle$ and $\langle \psi| H |\psi \rangle$ are denoted by $l^{H^\dag\! H}/r^{H^\dag\! H}$ and $l^{H}/r^{H}$.

$s_n$ will be close to $0$ when approaching convergence, a tiny error of $s_n$ in Eq.~(\ref{equ:dsde}) can lead to a significant error in the gradient.
To avoid this issue, we set the gradient manually to $g = \varepsilon - V_n^\dag HV_n$ and control the gradient descent with an adaptive learning rate $\alpha$, i.e, $\varepsilon \rightarrow \varepsilon-\alpha g$.
The learning rate is updated in each iteration.
Whenever the real part of $g$ changes sign, $\alpha$ is shrunk to $0.5 \alpha$ to ensure convergence, otherwise $\alpha$ is enlarged to $1.1 \alpha$ to accelerate convergence.
The pseudocode of GVMPS is shown in Algorithm~\ref{alg:GVMPS}.

\begin{algorithm}[tbp]
\DontPrintSemicolon
\KwInput{$N$-site non-Hermitian MPO $W$, $tol$, $max\_sweep$}
\KwOutput{ground state $|\rm SR \rangle$ MPS $\psi$}

get the ground energy $\varepsilon$ of $(H+H^\dag)/2$ using VMPS\;
generate random MPS $\psi$\;
initialize left and right environments:\;
\nonl\Indp $l^H, r^H$ = env($W$, $\psi$)\;
\nonl $l^{H^\dag\! H}, r^{H^\dag\! H}$ = env($W^\dag W$, $\psi$)\;
\Indm$\alpha = 1,\,g_0=0$\;

\For{$count = 1, \cdots, max\_sweep$}
{
\For($\qquad \qquad \triangleright \textrm{ sweep forward}$){$i = 1, \cdots, N-1$}
{
$W^{(2)}$ = $(W_i^\dag -\varepsilon^*)(W_i-\varepsilon)$\;
$H_{\rm eff}^{(2)}$ = getHeff($W^{(2)}, l^{H^\dag\! H}_i, r^{H^\dag\! H}_i$)\;
$\eta, \psi_i$ = eigs($H_{\rm eff}^{(2)}$, which = SR)\;
$H_{\rm eff}^{(1)}$ = getHeff($W_i, l^H_i, r^H_i$)\;
$e_i=\psi_i^\dag H_{\rm eff}^{(1)}\psi_i$\;
$\psi_i, U_R$ = QRdecomposition($\psi_i$)\;
$\psi_{i+1}$ = contraction($U_R, \psi_{i+1}$)\;
$l^H_{i+1}$ = updateLeftEnv($l^H_i, W_i, \psi_i$)\;
$l^{H^\dag\!H}_{i+1}$ = updateLeftEnv($l^{H^\dag\! H}_i, {W_i}^\dag W_i, \psi_i$)\;
}
$e_{\rm f}={\rm average}(e_i)$\;

\For($\qquad \qquad \triangleright \textrm{ sweep backward}$){$i =  N, \cdots, 2$}
{
$W^{(2)}$ = $(W_i^\dag -\varepsilon^*)(W_i-\varepsilon)$\;
$H_{\rm eff}^{(2)}$ = getHeff($W^{(2)}, l^{H^\dag\! H}_i, r^{H^\dag\! H}_i$)\;
$\eta, \psi_i$ = eigs($H_{\rm eff}^{(2)}$, which = SR)\;
$H_{\rm eff}^{(1)}$ = getHeff($W_i, l^H_i, r^H_i$)\;
$e_i=\psi_i^\dag H_{\rm eff}^{(1)}\psi_i$\;
$U_L, \psi_i$ = LQdecomposition($\psi_i$)\;
$\psi_{i-1}$ = contraction($\psi_{i-1}, U_L$)\;
$r^H_{i-1}$ = updateRightEnv($r^H_i, W_i, \psi_i$)\;
$r^{H^\dag\!H}_{i-1}$ = updateRightEnv($r^{H^\dag\! H}_i, {W_i}^\dag W_i, \psi_i$)\;
}
$e_{\rm b}={\rm average}(e_i)$\;

\If{${\rm abs}(e_{\rm f}-e_{\rm b})<10 \textrm{ } tol$}{
\If{${\rm abs}(\varepsilon-e_{\rm b})<tol$}{break}
$g=\varepsilon-e_b$\;
\uIf {$\Re\{g\}\times \Re\{g_0\}<0$}{$\alpha = 0.5\alpha$}\Else{$\alpha = 1.1\alpha$}
$\varepsilon = \varepsilon - \alpha \times g$\;
$g_0=g$\;
}
}
\Return{$\psi$}
\caption{
GVMPS
\label{alg:GVMPS}
}
\end{algorithm}

\section{EE for the PBC nH-SSH model\label{sec:pbc}}

\begin{figure}[tbp]
\centering
\includegraphics[width=1.0\columnwidth]{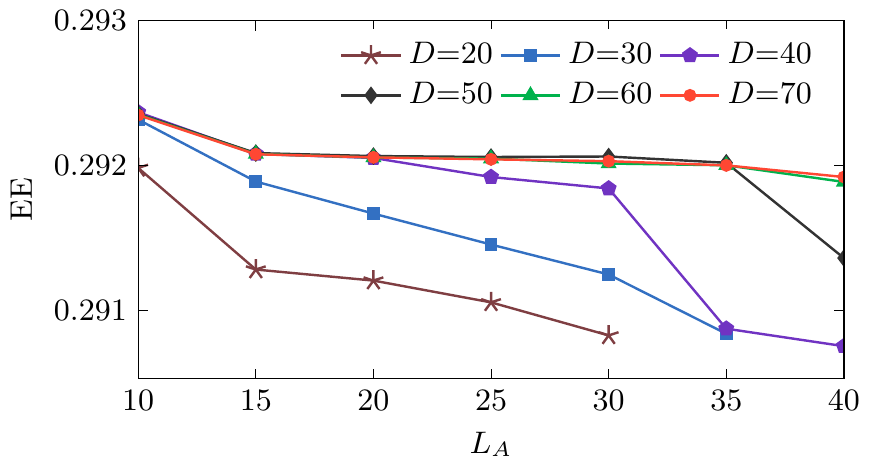}
\caption{
The maximum entanglement entropy of $|\rm SI\rangle$ calculated for the PBC nH-SSH model at $t=1$ and $\gamma=5$. 
EE converges at virtual bond dimension $D\ge 60$.}
\label{fig:svsl}
\end{figure}

In Fig.~\ref{fig:svsl}, we show the maximum entanglement entropy (EE) of $|\rm SI\rangle$ for the nH-SSH model of different system sizes at $t = 1$ and $\gamma = 5$.
EE exhibits asymptotic behavior with a plateau when the virtual bond dimension $D$ exceeds a certain threshold value.
This is consistent with the expected area law in the parameter region~\cite{Guo2021}, although the definition of EE is different.
Here we use MPS with OBC to simulate the PBC Hamiltonian for simplicity.

\section{Many-body particle distributions\label{sec:s1}}
To evaluate the many-body ground state distribution, we consider a general non-Hermitian single-particle model $H = \vec{a^\dag}M\vec{a}$ with dimension $L$.
$H$ can be diagonalized as $H=\vec{a^\dag}S\Lambda S^{-1}\vec{a}\triangleq\vec{d^\dag} \Lambda \vec{b}$, where $\Lambda={\rm Diag}(\{\lambda_i\})$, $\Re\{\lambda_i\}\le \Re\{\lambda_j\}$ for $i < j$, and
\begin{equation}
\begin{array}{lll}
\{b_i,b_j\}=0, & \{b^\dag_i,b^\dag_j\}=0, & \{b_i,b^\dag_j\}\ne\delta_{ij}; \\
\{d_i,d_j\}=0, & \{d^\dag_i,d^\dag_j\}=0, & \{d_i,d^\dag_j\}\ne\delta_{ij}; \\
\{d_i,b_j\}=0, & \{d^\dag_i,b^\dag_j\}=0, & \{b_i,d^\dag_j\}=\delta_{ij}.
\end{array}
\label{equ:com}
\end{equation}
Although neither $b_i$ nor $d_i$ satisfies the fermion anti-commutation relations, together they form a set of bi-fermionic operators~\cite{Chang2020,Alsallom2021}. 

For nonredundant indices in ascending order $\{i_j| 1\le j\le p\}$, $d^\dag_{i_1} d^\dag_{i_2}\cdots d^\dag_{i_p}|0\rangle$ is an $p$-particle right eigenvector of $H$ with energy $\sum_{1\le j\le p}\lambda_{i_{j}}$, and the corresponding left eigenvector is $\langle 0|b_{i_p} b_{i_{p-1}}\cdots b_{i_1}$. 
The particle distribution of the $p$-particle state $|\mathrm{SR}\rangle = d^\dag_1 d^\dag_2 \cdots  d^\dag_p|0\rangle$ on the $m$-th site is given by
\begin{equation}
\begin{aligned}
\langle n_{m}\rangle_p = & \frac{\left|a_m d^\dag_1 d^\dag_2 \cdots  d^\dag_p|0\rangle\right|^2}{\left|d^\dag_1 d^\dag_2 \cdots d^\dag_p |0\rangle\right|^2} \\
= & \frac{\left|\sum\limits_{\{q_{(L)}^{(p)}\}} a_m \prod\limits_{1\le i\le p} S_{q_i,i} a^\dag_{q_i}|0\rangle\right|^2}{\left|\sum\limits_{\{q_{(L)}^{(p)}\}} \prod\limits_{1\le i\le p} S_{q_i,i} a^\dag_{q_i}|0\rangle\right|^2}.
\end{aligned}
\label{equ:npm}
\end{equation}
Here $\{q_{(L)}^{(p)}\}$ represents one of the $p$-permutations of the set $\{1,2,\cdots,L\}$, the total number of which is $L!/(L-p)!$.
We rearrange the order of creation operators using the fermion anti-commutation relations.
\begin{equation}
\prod_{1\le i\le p} a_{q_i}^\dag |0\rangle=| \{q_{(L)}^{(p)}\} \rangle =\epsilon_{\{q_{(L)}^{(p)}\}}|[q_{(L)}^{(p)}] \rangle,
\end{equation}
where $[q_{(L)}^{(p)}]$ denotes one of the $p$-combinations of the set $\{1,2,\cdots,L\}$ with $q_1<q_2<\cdots<q_p$, the total number of which is $\frac{L!}{p!(L-p)!}$.
$\epsilon_{\{q_{(L)}^{(p)}\}}$ denotes
\begin{equation}
\epsilon_{\{q_{(L)}^{(p)}\}}=\begin{cases}
+1 & \{q_{(L)}^{(p)}\}\mathrm{\textrm{ is an even permutation of }}[q_{(L)}^{(p)}],\\
-1 & \{q_{(L)}^{(p)}\}\mathrm{\textrm{ is an odd permutation of }}[q_{(L)}^{(p)}],\\
0 & \textrm{otherwise}.
\end{cases}
\end{equation}
When combined with $\prod\limits_{1\le i\le p} S_{q_i, i}$ and summed over all terms contributing to the same state, the coefficient in front of $|[q_{(L)}^{(p)}]\rangle$ is the determinant of a sub-matrix of $S$ selected from row indices $[q_{(L)}^{(p)}]$ and column indices $(1,2,\cdots,p)$, denoted as $S_{[q_{(L)}^{(p)}],[p]}$, 
\begin{equation}
\sum\limits_{\{q_{(L)}^{(p)}\}} \prod\limits_{1\le i\le p} S_{q_i,i} a^\dag_{q_i}|0\rangle= \sum_{[q_{(L)}^{(p)}]} \det(S_{[q_{(L)}^{(p)}],[p]})| [q_{(L)}^{(p)}] \rangle.
\label{equ:det}
\end{equation}

Now we consider the extra annihilation operator $a_m$ in the numerator of Eq.~(\ref{equ:npm}). 
It is clear that only terms with exactly one $q_j = m$ contribute to the summation.
Using the fermion anti-commutation relation, we have
\begin{equation}
a_m \prod\limits_{1\le i\le p} a^\dag_{q_i}|0\rangle = {(-1)}^{(j-1)}\prod\limits_{q_i\ne m} a^\dag_{q_i}|0\rangle.
\end{equation}
Thus, similar to Eq.~(\ref{equ:det}),
\begin{equation}
\begin{aligned}
& \sum\limits_{\{q^{(p)}\}} a_m \prod\limits_{1\le i\le p} S_{q_i,i} a^\dag_{q_i}|0\rangle \\
= & \sum_{[q_{(L|\bar{m})}^{(p-1)}]} \det(S_{(m,[q_{(L|\bar{m})}^{(p-1)}]),[p]})|[q_{(L|\bar{m})}^{(p-1)}]\rangle \\
= & \sum_{[q_{(L|\bar{m})}^{(p-1)}]} \sum_{j=1}^{p} (-1)^{j-1} S_{m,j} \det(S_{[q_{(L|\bar{m})}^{(p-1)}],[p|\bar{j}]}) |[q_{(L|\bar{m})}^{(p-1)}]\rangle.
\end{aligned}
\end{equation}
Here, $[q_{(L|\bar{m})}^{(p-1)}]$ denotes one of the $(p-1)$-combinations of the set $\{1,2,\cdots,m-1,m+1,\cdots,L\}$ with $q_1 < q_2<\cdots<q_{p-1}$, the total number of which is $\frac{(L-1)!}{(p-1)!(L-p)!}$.
$[p|\bar{j}]$ denotes $(1,2,\cdots,j-1,j+1,\cdots,p)$.

The squared norm in Eq.~(\ref{equ:npm}) involves a multiplication of two determinants, similar to the Cauchy-Binet formula
\begin{equation}
\begin{aligned}
\det(AB) = & \det(A_{[M],[N]}B_{[N],[M]}) \\
= & \sum_{[q_{(N)}^{(M)}]} \det(A_{[M],[q_{(N)}^{(M)}]})\det(B_{[q_{(N)}^{(M)}],[M]}),
\end{aligned}
\end{equation} 
where $A$ and $B$ are $M\times N$ and $N\times M$ matrices with $N>M$.
Using this formula from right to left, Eq.~(\ref{equ:npm}) is simplified to
\begin{equation}
\begin{aligned}
& \langle n_{m} \rangle_p \\
= & \frac{\sum_{i,j=1}^{p} (-1)^{i+j} S^{\dagger}_{i,m} S_{m,j} \det(S^\dag_{[p|\bar{i}],[L|\bar{m}]}S_{[L|\bar{m}], [p|\bar{j}]})}{\det(S^\dag_{[p],[L]}S_{[L],[p]})}.
\end{aligned}
\label{equ:result}
\end{equation}
To calculate bi-orthogonal particle distributions constructed by $b$ and $d^{\dagger}$ operators, we simply replace all $S^\dag$ in Eq.~(\ref{equ:result}) with $S^{-1}$.

The results of Eq.~(\ref{equ:result}) for small systems agree with those from exact diagonalizations. 
Direct calculations from Eq.~(\ref{equ:npm}) require more than $O(L^p)$ steps, where $L$ is the system size. 
Using the Cauchy-Binet formula, $\langle n_{m}\rangle_p$ can be calculated in at most $O(Lp^4)$ steps. 
To prevent loss of precision for large systems~\cite{Goldberg1991}, all variables in Eq.~(\ref{equ:result}) are converted to floating-point numbers with high precision.
The remarkable agreement between single-particle results and HVMPS results confirms the validity of our method.

\section{Symmetry of particle distributions}

The Hamiltonian of the $N$-unit-cell nH-SSH model is $H=\vec{a^\dag}M\vec{a}$, which can be diagonalized as $H=\vec{a^\dag}S\Lambda S^{-1}\vec{a}\triangleq\vec{d^\dag}\Lambda\vec{b}$.
The single particle Hamiltonian $M$ is a $2N\times2N$ matrix
\begin{equation}
\left(\begin{array}{ccccccc}
    0          & t+\gamma' &            &            &        &            &            \\
    t-\gamma' & 0          & 1          &            &        &            &            \\
     & 1          & 0          & t+\gamma' &        &            &            \\
     &            & t-\gamma' & 0          & 1      &            &            \\
     &            &            & 1          & 0      & \ddots     &            \\
     &            &            &            & \ddots & \ddots     & t+\gamma' \\
     &            &            &            &        & t-\gamma' & 0
\end{array}\right)
\end{equation}
with $\gamma'=\gamma/2$.
For the diagonal matrix ${\rm Diag}(X)=\{1,-1,\cdots,1,-1\}$, we find $-M=XMX=(XS)\Lambda (S^{-1}X)$. 
Thus, the eigenvalues $\{\lambda_k|k \in [2N]\}$ and the corresponding right (left) eigenvectors $\{S_{[2N],k}\}$ ($\{{(S^{-1})}_{k,[2N]}\}$) satisfy
\begin{equation}
\begin{aligned}
\lambda_k        & = -\lambda_{2N-k+1}       \\
S_{[2N],k}          & = X S_{[2N],2N-k+1}          \\
{(S^{-1})}_{k,[2N]} & = {(S^{-1})}_{2N-k+1,[2N]}X.
\end{aligned}\label{equ:x}
\end{equation} 
Multiplying by $X$ does not change particle distributions, so the single-particle distributions are symmetric for states with energies $\lambda_{k}$ and $\lambda_{2N-k+1}$. 

In Sec.~\ref{sec:s1}, we construct a $p$-particle state from the vacuum state $|0\rangle$, where $p$ lowest energy modes are filled. 
When there are more than half-filled particles, it is more convenient to start from the full-filled state $|2N\rangle$.
A direct derivation gives
\begin{equation}
\begin{aligned}
& d_1^\dag\cdots d_p^\dag|0\rangle \\
= & b_{2N}\cdots b_{p+1}d_1^\dag\cdots d_{2N}^\dag|0\rangle \\
= & b_{2N}\cdots b_{p+1}|2N\rangle \\
= & \sum_{\{ q_{(2N)}^{(2N-p)} \}} \prod\limits_{1\le i\le 2N-p}{(S^{-1})}_{2N-i+1,q_i} a_{q_i}|2N\rangle \\
= & \sum_{\{ q_{(2N)}^{(2N-p)} \}} \prod\limits_{1\le i\le 2N-p}{(S^{-1})}_{i,q_i}X_{q_i q_i} a_{q_i}|2N\rangle.
\end{aligned}
\end{equation}
On the other hand,
\begin{equation}
\begin{aligned}
& b_{1}\cdots b_{2N-p}|2N\rangle = \sum_{\{ q_{(2N)}^{(2N-p)} \}} \prod\limits_{1\le i\le 2N-p}{(S^{-1})}_{i,q_i} a_{q_i}|2N\rangle.
\end{aligned}
\end{equation}
The $X_{q_i q_i}$ term can be absorbed into $a_{q_i}$ without changing particle distributions. 
Therefore, $d_1^\dag\cdots d_p^\dag|0\rangle$ can be replaced by $b_1\cdots b_{2N-p}|2N\rangle$ in subsequent calculations.
\begin{equation}
\begin{aligned}
\langle n_{m}\rangle_{2N-p}
= & \frac{\langle0| d_{2N-p}\cdots d_1 a_m^\dag a_m d_1^\dag\cdots d_{2N-p}^\dag|0\rangle}{\langle0| d_{2N-p}\cdots d_1 d_1^\dag\cdots d_{2N-p}^\dag|0\rangle} \\
= & \frac{\langle 2N| b^\dag_p\cdots b^\dag_1 a_m^\dag a_m b_1\cdots b_p|2N\rangle}{\langle 2N| b^\dag_p\cdots b^\dag_1  b_1\cdots b_p|2N\rangle} \\
= & 1-\frac{\langle 2N| b^\dag_p\cdots b^\dag_1 a_m a_m^\dag b_1\cdots b_p |2N\rangle}{\langle 2N| b^\dag_p\cdots b^\dag_1  b_1\cdots b_p|2N\rangle}.
\end{aligned}
\end{equation}
The second term in the above equation is evaluated in the same way as the particle distributions of the left eigenvector $\langle \rm SR|$.
As mentioned in Appendix~\ref{sec:bi}, left eigenvectors of nH-SSH model are spatial reflections of right eigenvectors. 
Thus, $1-\langle n_{m}\rangle_{2N-p}$ is the spatial reflection of $\langle n_{m}\rangle_p$.
If we consider the particle distribution with respect to the unit-cell index $i$ and each unit cell has two sites, the above relation becomes $\langle n_{i} \rangle_p = 2- \langle n_{N+1-i} \rangle_{2N-p}$.

\end{document}